\documentclass{aa}  

\usepackage{graphicx}
\usepackage{color}
\usepackage{txfonts}
\usepackage{longtable}

\def\ergs{\ifmmode erg s^{-1} \else erg s$^{-1}$\fi}

\def\msun{\ifmmode M_{\odot} \else M$_{\odot}$\fi}
\def\msunyr{\ifmmode M_{\odot} {\rm yr}^{-1} \else M$_{\odot}$ yr$^{-1}$\fi}
\def\zsun{\ifmmode Z_{\odot} \else Z$_{\odot}$\fi}
\def\lsun{\ifmmode L_{\odot} \else L$_{\odot}$\fi}

\newcommand{\oh}{\ifmmode 12 + \log({\rm O/H}) \else$12 + \log({\rm
O/H})$\fi}
\newcommand{\hii}{H~{\sc ii}}
\newcommand{\heii}{He~{\sc ii}}
\newcommand{\nev}{[Ne~{\sc v}]}
\newcommand{\hep}{He$^+$}
\newcommand{\nii}{[N~{\sc ii}]}
\newcommand{\oi}{[O~{\sc i}]}
\newcommand{\oii}{[O~{\sc ii}]}
\newcommand{\oiii}{[O~{\sc iii}]}
\newcommand{\ha}{\ifmmode {\rm H}\alpha \else H$\alpha$\fi}
\newcommand{\hb}{\ifmmode {\rm H}\beta \else H$\beta$\fi}
\newcommand{\Nii}{[N~{\sc ii}] $\lambda\lambda$6548,6584}

\newcommand{\Oi}{[O~{\sc i}] $\lambda$6300}
\newcommand{\Oii}{[O~{\sc ii}] $\lambda$3727}
\newcommand{\Oiii}{[O~{\sc iii}] $\lambda\lambda$4959,5007}
\newcommand{\oiiil}{[O~{\sc iii}]$\lambda 5007$}
\newcommand{\Heii}{He~{\sc ii}$\lambda$4686}
\newcommand{\Heiiuv}{He~{\sc ii}$\lambda$1640}
\newcommand{\Nev}{[Ne~{\sc v}]$\lambda$3426}
\newcommand{\Civ}{C~{\sc iv} $\lambda$1550}

\newcommand{\Ciiiuv}{C~{\sc iii}] $\lambda$1909}

\newcommand{\lxlong}{\ifmmode L_{X,0.5-8 {\rm keV}} \else $L_{X,0.5-8 {\rm keV}}$\fi}  
\newcommand{\lx}{\ifmmode L_X \else $L_X$\fi}

\begin{document}

\title{Can nebular \heii\ emission be explained by ultra-luminous X-ray sources ?}

   \author{Charlotte Simmonds\inst{1} 
   \and Daniel Schaerer\inst{1,2}
   \and Anne Verhamme\inst{1,3}
   }

\institute{
    Observatoire de Gen\`eve, D\'epartement d'Astronomie, Universit\'e de Gen\`eve, 51 Chemin Pegasi, 1290 Versoix, Switzerland
    \and
    CNRS, IRAP, 14 Avenue E. Belin, 31400 Toulouse, France
    \and    
    Univ. Lyon, Univ. Lyon1, Ens de Lyon, CNRS, Centre de Recherche Astrophysique de Lyon UMR5574, F-69230, Saint-Genis-Laval, France
    }

\authorrunning{Simmonds, Schaerer \& Verhamme}
\titlerunning{Nebular \heii\ emission from HMXB/ULX}


  \abstract
   {The shape of the ionising spectra of galaxies is a key ingredient to reveal their physical properties and to our understanding of the ionising background radiation. A long-standing unsolved problem is the presence of \heii\ nebular emission in many low-metallicity star-forming galaxies. 
   This emission requires ionising photons with energy $>54$ eV, which are not produced in sufficient amounts by normal stellar populations.}
   {To examine if high mass X-ray binaries and ultra-luminous X-ray sources (HMXB/ULX) can explain the observed \heii\ nebular emission and how their presence alters other emission lines, we compute photoionisation models of galaxies including such sources.}
   {We combine spectral energy distributions (SEDs) of integrated stellar populations with constrained SEDs of ULXs to obtain composite spectra with varying amounts of X-ray luminosity, parameterised by $L_X/$SFR. With these we compute photoionisation models to predict the emission line fluxes of the optical recombination lines of H and He+, and the main metal lines of \oiii, \oii, \oi, and \nii. The predictions are then compared to a large sample of low-metallicity galaxies.}
   {We find that it is possible to reproduce the nebular \Heii\ and other line observations with our spectra and with amounts of $L_X/$SFR compatible with the observations.
   Our work suggests that HMBX/ULX could be responsible for the observed nebular \heii\ emission. However, the strengths of the high and low ionisation lines, such as \heii\ and \Oi, depend strongly on the X-ray contribution and on the assumed SEDs of the high energy source(s); the latter are poorly known.}
   {}

\maketitle
%

\section{Introduction}

The presence of strong emission lines and the detection of intense UV metal lines in numerous high-redshift galaxies, and more generally the quest for the sources of cosmic reionisation, has re-triggered a strong interest in the physical properties of distant galaxies and their radiation field \citep[see e.g. the review of ][]{Stark2016}. In particular, prominent high ionisation spectral features, such as nebular emission in \Civ, \Heiiuv, as well as intense \Ciiiuv\ emission, have for example been detected in several $z>6$ galaxies, raising thus questions about the origin of their hard ionising spectra \citep{Stark2015,Schmidt2017} Indeed, galaxies showing these UV emission lines are relatively rare at intermediate redshifts \citep[$z \sim 2-3$, see e.g.][]{Erb2010,LeFevre2019,Schmidt2021} and in the low-$z$ Universe \citep[see e.g.][]{Senchyna2017,Berg2019}.

Low-$z$ galaxies showing nebular emission in \Civ, \heii\ in the UV (\Heiiuv) and optical (\Heii), and other high ionisation lines, such as \Nev\ are known, primarily among metal-poor star-forming galaxies \citep{Berg2019,Senchyna2017,Guseva2000,Shirazi2012}. In the IR domain, the fine structure lines of [O~{\sc iv}] 25.9 $\mu$m, [Ne~{\sc v}] 14.3$\mu$m and 24.3$\mu$m, for example, are also detected in certain star-forming galaxies and their strength increases in active galactic nuclei (AGN) \citep{Sturm2002,Fernandez2016}. The relative weakness and paucity of these lines in star-forming galaxies is not surprising, since nebular \Civ, \Heii, O~{\sc iv}, and Ne~{\sc v} emission lines require photons with energies above 47.9, 54.4, 54.9, and 97.1 eV respectively, which are relatively sparsely emitted by stars and stellar populations and thus thought to be associated with other sources of higher energy photons.

In fact, the presence and strength of nebular \Heii\ in optical spectra (and that of [O~{\sc iv}] in IR spectra) of star-forming galaxies has been recognised already early on as a puzzle/problem \citep[e.g.][]{Schaerer1996,Guseva2000,Shirazi2012}, except in galaxies showing significant populations of Wolf-Rayet (WR) stars, which could be hot and numerous enough to produce sufficient doubly ionised He (He$^{2+}$) \citep[e.g.][]{Schaerer1996,Schaerer1999}. ``Normal'' stellar populations do not predict sufficiently hard ionising spectra above 54 eV.
Given that WR stars become much rarer at low metallicity (\oh $\la 8.2$), other sources or processes must be found to explain nebular \heii\ emission in these galaxies. Furthermore, nebular \Heii\ is quite ubiquitous at these low metallicities\footnote{ For example, in the sample analysed by \cite{Schaerer2019} \heii\ is detected in $\sim 2/3$ of the galaxies. In the larger sample analysed here this is the case for $\sim 1/3$ of the star-forming galaxies.}, 
and the average observed intensity \heii/\hb\ increases with decreasing O/H. This describes the essence of the nebular \heii\ problem.

Despite constant improvements in stellar evolution and atmosphere models this problem has not yet been solved. Indeed, the nebular \heii\ problem has by large been confirmed by many more recent works using different up-to-date stellar population and photoionisation models, and both analysing the optical and UV recombination lines of \Heii\ and \Heiiuv\ from different galaxy samples \citep[e.g.][]{Stasinska2015,Steidel2016,Nanayakkara2019,Saxena2020AA,Stanway2019,Plat_2019,Berg2021}.

On the other hand, various alternate sources and mechanisms have been explored to attempt solving the nebular \heii\ problem.
For example, radiative shocks have been proposed to explain (both) the observed \Heii\ emission (and \Nev) in some low-metallicity galaxies
\citep[see e.g.][]{Thuan2005,Izotov2012,Plat_2019}.
X-rays, from sources such as high-mass X-ray binaries (HMXB) and ultra-luminous X-ray sources (ULX), have been advocated as the origin of nebular \heii\ emission. \citep[e.g.][]{Schaerer2019,Oskinova2019}.
Other authors argued for possible contributions of Pop III-like (metal-free) stars, fast-rotating stars, stripped stars in binary systems or similar, which could boost the emission of \heii\ ionising photons \citep[see][]{Cassata2013,Szecsi2015,Gotberg2019,Stanway2019}.
Recently, some studies have also proposed ``unconventional'' explanations, which do not require additional ionising sources or peculiar stars
\citep{Barrow2019,PerezMontero2020},  and which are discussed later (Sect.\ \ref{sect_other}).
Clearly, no consensus has yet been achieved on the origin of nebular \heii, and each of the above scenarios can be debated.
Furthermore, several of them do not make any quantitative predictions which can be compared with observations and tested.

In the present paper, we examine the impact of X-ray sources (HMBX, ULX) on nebular \heii\ and other emission lines in a quantitative manner, using Cloudy photoionisation models. 
Our approach is motivated by the simple model of \cite{Schaerer2019}, who compared the observed increase of the X-ray luminosity per unit star-formation rate ($L_X/$SFR) with decreasing metallicity \citep{Brorby2016,Douna2015} with the observed increase of the \heii/\hb\ intensity, and proposed a causal connection between the two. Such a connection appears quite natural since the X-ray emission in star-forming galaxies is dominated by individual bright sources \citep[HMXB and/or ULX,][]{Mineo2012b} and several ULXs are known to show nebulae with \heii\ emission and also [O~{\sc iv}] 25.9 $\mu$m \citep{Pakull1986,Pakull2003,Kaaret2009,Berghea2012}.
Postulating a simple scaling between the He$^+$ ionising and X-ray flux taking from observations, \cite{Schaerer2019} have shown that the observed \heii/\hb\ intensities and their metallicity dependence can approximately be explained.

Several studies have criticised the recent HMXB/ULX scenario.
For example, \cite{Kehrig2021} argued that the observations of the low metallicity galaxy I Zw 18 used by \cite{Schaerer2019} for the scaling, show an insufficient contribution of the ULX in this galaxy to \heii\ ionising photons. New integral field spectroscopic observations with the Keck telescope of I Zw 18 provide a less definite answer on this issue, 
and \cite{Rickards2021} argue that shocks or beamed X-ray emission from ULXs could be responsible for the \heii\ emission.
\cite{Senchyna2020} report the absence of bright X-ray sources in several nearby galaxies with nebular \heii.
Furthermore, using simple photoionisation models combining HMXBs described by a multi-colour disc model with normal stellar populations, \cite{Senchyna2020} have concluded that HMXBs are inefficient producers of the photons necessary to power \heii.

On the other hand, observations of ULX in nearby galaxies have often revealed the presence of nebulae, some, but not all of them, showing high ionisation emission lines
\citep[cf.\ review of][]{Kaaret2017}. 
For example, nebular \Heii, \Nev, and [O~{\sc iv}] 25.9 $\mu$m emission has been observed by \cite{Pakull2003,Lehmann2005,Abolmasov2008,Kaaret2009,Berghea2010,Grise2011,Maggi2011,Moon2011Large-Highly-Io,Berghea2012,Binder2018,Urquhart2018,Vinokurov2018} 
in different ULX nebulae, including famous objects such as Holmberg II X-1, NGC5408 X-1, and others.
Similarly, nebular \heii\ is also seen in HMXB of lower X-ray luminosity \citep[e.g.][]{Pakull1986}.
Observationally, HMXB/ULX can thus produce ionising photons which give rise to some or several of the high ionisation lines discussed above and observed in low-metallicity star-forming galaxies. 
In addition HMXB/ULX are known and predicted to be more numerous/luminous at low metallicities, their contribution is expected to be dominant in distant star-forming galaxies, and $L_X/$SFR increases towards high redshift \citep{Fragos2013,Lehmer2016,Madau2017,Fornasini2019,Kovlakas2020}. 
It is therefore of interest to explore the effect of ULXs whose spectra are constrained by observations, on integrated galaxy spectra.

For this purpose we here construct spectral energy distributions (SEDs) of star-forming galaxies including a contribution from ULXs. The stellar spectra, taken from the BPASS binary population synthesis code \citep{Eldridge2017}, are combined with three SEDs of ULX sources with nebulae, and whose spectra have been constrained by multi-wavelength observations by
\cite{Kaaret2009} and \cite{Berghea2012}.
Combined SEDs are constructed allowing for variable amounts of X-ray luminosities, corresponding to the observed range of $L_X/$SFR in star-forming galaxies.
With these SEDs we compute small grids of Cloudy photoionisation models to predict the most relevant optical emission lines, which are then compared to observations of a large sample of star-forming galaxies for which measurements of nebular \heii\ are available.

Our methodology is described in Sect.\ \ref{sect_models}. In Sect.\ \ref{sect_results} we present the predicted high ionisation emission line strengths as well as the most important optical emission lines using various emission line diagrams, showing the impact of varying amounts of X-ray emission on these observables.
We discuss our results and other recent approaches in Sect.\ \ref{sect_discuss}. Our main conclusions are summarised in Sect.\ \ref{sect_conclude}.

\begin{figure}
   \centering
   \includegraphics[width=\columnwidth]{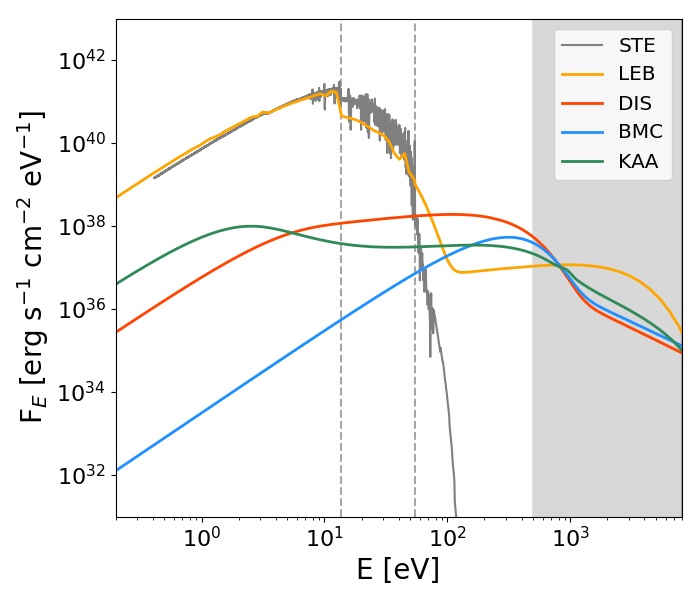}
   \caption{Basic SEDs used for the photoionisation models: 
  the three ULX spectra (DIS, BMC, KAA), normalised to the same X-ray luminosity \lx, are shown by the red, blue, and green lines respectively.
  For comparison the pure stellar SED, labelled STE and shown by the grey line, is scaled to our standard value of $\lx/{\rm SFR}=7.9\times10^{39}$ \ergs/(\msunyr); see text.
  The SED of I Zw 18 (LEB model) from \cite{Lebouteiller2017}, is shown in yellow, scaled to the same stellar UV flux at $\sim 10$ eV as STE, resulting in an \lx\ 
   higher by a factor $\sim 2.8$ compared to the ULX spectra. 
  The vertical dashed lines mark the ionisation potentials of hydrogen and helium (13.6 eV and 54 eV, respectively). The grey shaded area highlights the 0.5-8 keV range of the spectra used for $L_X$.}
  \label{fig_norms}
\end{figure}

\section{Photoionisation models of star-forming galaxies including ULXs}
\label{sect_models}

To predict the emission line spectra of star-forming galaxies which harbour HMXB or ULX sources, we need to compute detailed photoionisation models using appropriate
spectral energy distributions. 
Since generally the available, theoretical SEDs of galaxies do not include X-ray sources, we first compile several spectra of ULXs from the literature, which we then combine 
-- using an appropriate scaling -- with SEDs from evolutionary synthesis models describing the ``normal'' stellar populations, i.e.\ excluding the contribution from ULXs.
We now briefly describe the SEDs we use, the way they are combined, and the Cloudy model grid we have computed.

\subsection{SEDs of ULXs}

The SEDs of ULXs are relatively poorly known, and show a large diversity and also variability \citep[e.g.][]{Gladstone2009,Sutton2013a,Sutton2013b,Walton2015,Pintore2016,Gurpide2021b}.
   In general, the X-ray spectra of ULXs is modelled by the sum of two components related to the emission from the accretion flow produced close to the compact object: a multi-colour black body for the accretion disk and a power law representing the coronal emission \citep{Kaaret2017}. 
Also, the UV/optical spectra can often be fitted by a companion star or an irradiated (sometimes supercritical)  disk \citep[cf.][]{Grise2012,Tao2012,Gurpide2021a}, which further illustrates the uncertainties in the SEDs.

From the literature we have compiled three SEDs of ULXs, which have been observed not only at X-ray wavelengths but also at longer wavelengths (optical, some also in the infrared), and for which tailored photoionisation models fitting the X-ray data and other observational constraints have been constructed. These spectra cover the entire spectral range from the X-rays to the extreme-UV, extending down to low energies ($ \sim 54$ eV and lower), which is mandatory to allow combinations with stellar spectra and to explore in particular their impact on \heii\ lines, whose ionisation potential is 54 eV. In particular, throughout this work, we assume that the radiation in the UV-optical regimes is predominantly produced by stars, while the radiation in the X-rays is produced by the ULX. We comment later on the choice and representativeness of these SEDs (Sect.\ \ref{s_diversity}).

The SEDs used in this paper are shown in Fig.~\ref{fig_norms}, and their main relevant properties, to be discussed below, are summarised in Table \ref{table_ulx_models}. 

In short, we list two indicators of spectral hardness: the first gives the ratio of the ionising photon fluxes above 54 eV, $Q_2$, over those above 13.6 eV, $Q_0$, which measures the hardness of the ionising spectrum above the \hep\ potential to that of hydrogen. The second quantity, $Q_2/L_X$ measures the relative number of \hep\ ionising photons emitted per X-ray luminosity (0.5-8 keV in this work), i.e.\ an efficiency of the X-ray source to produce \heii\ recombination line emission. It is listed only for SEDs with an X-ray component. $Q_0/Q_2$ is not given for pure ULX SEDs, since their contribution to H-ionisation is negligible in combined models. 

\begin{table}[tb]
{
\hfill{}
   \begin{center}
   \caption[]{Spectral properties of the basic ULX and stellar SEDs} 
       \begin{tabular}{lcc}
                \hline
                \noalign{\smallskip}
                Model & log($Q_0$/$Q_2$) & log($Q_2/\lx$) [photon/erg] \\
            \noalign{\smallskip}
                \hline
                \noalign{\smallskip}
                        STE  &  4.04  &  - \\ 
                        LEB  &  2.19  &  9.50 \\ 
                        DIS  &  -  &  10.25 \\ 
                        BMC  &  -  &  9.65 \\ 
                        KAA  &  -  &  9.61 \\
                        \noalign{\smallskip}
                \hline
        \end{tabular}
     \tablefoot{{\it Col. 1:} model name. The stellar model (STE) refers to a metallicity of 12 + log(O/H) = 8.1.
     {\it Col. 2:} ratio between $Q_0$ and $Q_2$. {\it Col. 3:} ratio between $Q_2$ and the 0.5-8 keV X-ray luminosity.
     Empty columns (-) indicate when the SEDs are not or not well defined in one of the spectral ranges considered.}
\label{table_ulx_models}
   \end{center}
   }
   \end{table}
   
\subsubsection{NGC 6946 X-1 SEDs}
   Two SEDs are taken from \cite{Berghea2012}, who have produced a model for the full optical to X-ray SED of the ULX NGC 6946 X-1 associated to the MF 16 Nebula. 
   This ULX has very similar properties to the famous Holmberg II ULX \citep[e.g.][]{Miller2005,Goad2006,Walton2015} and emission in the infrared nebular [O~{\sc iv}] line, often seen in active galactic nuclei, was detected by \cite{Berghea2010}.
   Several X-ray models were used to reproduce published \textit{Hubble} optical/UV data along with data obtained via X-ray spectroscopy. 

   \cite{Berghea2012} found two XSPEC \citep{Arnaud1996} Comptonisation models which can fit the X-ray data equally well.
   \begin{itemize}
       \item DISKIR (Irradiated Disk; hereafter DIS): this model represents an accretion disk whose outer region is irradiated by its inner region and the coronal emission of the black hole. A fraction of the bolometric flux of the inner disk is thermalised to the local blackbody temperature at each radius, changing the thermal structure of the outer regions and consequently, the shape of the SED. This model could fit a fraction of the ULXs and was originally presented as a representation of binary black hole systems. In it, the bulk of the disk's flux is reflected and only the hardest X-ray photons heat the disk \citep{Gierlinski2009}. 
       \item BMC (Comptonisation by Relativistic Matter): this is an analytic model that describes the
       Comptonisation of soft photons by matter undergoing relativistic bulk-motion. It is typically used to describe thermal X-rays which originate in the inner region of an accretion disk in a binary black hole system \citep{Gliozzi2011}, the in-falling matter is heated in the vicinity of the event horizon of the black hole, generating high energy photons \citep{Titarchuk1997}.
   \end{itemize}
   These models specifically describe high mass X-ray systems. 
   The difference between these models is found in the predictions they yield for the optical/UV data. Specifically, the BMC model requires a bright stellar companion in order to fit this part of the spectrum of the MF16 nebula, while the DIS model does not. We include the BMC model without stellar companion to our calculations because we are interested in the high-energy spectrum of the ULX, which will be combined with a stellar population model in order to produce a complete SED.
   For this work we use both models, BMC and DIS, to which we add stellar emission from the integrated population.
\

\subsubsection{NGC 5408 X-1 SED}

   The third SED (KAA, hereafter) is a model of a photoionised nebula created by analysing the spectra of the optical counterpart of the ULX NGC 5408 X-1 as described in \cite{Kaaret2009}. 
   Some authors argue that this source hosts an intermediate-mass black hole candidate, whereas others argue for a stellar- mass compact object accreting at super-Eddington rate \citep[cf.][]{Kaaret2009,Middleton2011}.
   The spectra of this nebula show high excitation emission lines, in particular \Heii\ and \Nev, which suggest X-ray photoionisation \citep{Kaaret2009,Cseh2012}.
   The KAA SED is based on a Comptonisation model \cite[\texttt{compps},][]{Poutanen1996} and includes a multi-colour blackbody representing the emission from the wind or the outer parts of the supercritical disk and a cutoff powerlaw as a proxy for the emission emitted closer to the compact object \citep{Gladstone2009}.
    This model works by using exact numerical solutions to the radiative transfer equation and includes parameters such as geometry, optical depth, electron distribution, spectral distribution of soft seed photons and angle of the observer.  

\subsection{A galaxy SED including an ULX -- I Zw 18}
    In addition to the above SEDs, which primarily focus on the X-ray part of the spectrum and describe ULXs, we have also used a SED of the well-known metal-poor galaxy I Zw 18, which harbours a ULX source in the centre of its main \hii\ region, I Zw 18-NW.  Detailed multi-wavelength observations covering numerous emission lines and tracers from different ISM phases are available for this region. \cite{Lebouteiller2017} have studied this object in depth and demonstrated the importance of the ULX for the understanding of this giant \hii\ region. Their SED for the northwestern region (standard, \lx\ = $4\times 10^{40}$ \ergs), including stellar and X-ray emission constructed using multiple, empirically adjusted components and an accretion disk model (diskbb) for the X-ray source, is shown in Fig.\ \ref{fig_norms} and denoted by LEB subsequently. 
    As discussed in \cite{Lebouteiller2017}, the SED was empirically adjusted at energies close the He$^+$ ionisation potential.
    Since this is a ``combined'' and empirically-constrained galaxy SED, covering the entire spectral range (from UV to X-rays) or relevance here, the X-ray contribution is kept fixed.

\subsection{Photoionisation modelling using combined SEDs}
\
Finally, to produce complete SEDs including stellar and X-ray emission, we combine the above ULX spectra with SEDs obtained from the binary population synthesis code BPASS v2.1 \cite{Eldridge2017}. We adopt the predicted SED of a young population (age $10^6$ yr), with a Kroupa IMF allowing for stellar masses up to 100 \msun, and a metallicity of 12 + log(O/H) = 8.1.
In this work we aim to study the effect of the X-ray contribution that ULXs can have over a galaxy, thus, we have focused on one stellar metallicity only, chosen to be representative of our observational sample (cf.\ below). Small metallicity variations in the stellar model do not produce significant changes in the SED shape or in the resulting emission line intensities.
The young age was chosen for comparison with earlier studies, which commonly use such ages for comparisons with observations (BPT diagrams and others), since this provides relatively
hard ionising spectra \citep[see e.g.][]{Kewley2006,Stasinska2015,Nakajima2018}.
The stellar SED is denoted as STE.

For the combination of ULX and stellar SEDs we are guided by the observed correlation between the X-ray luminosity and star-formation rate (SFR) in star-forming galaxies, which also shows a metallicity-dependence \citep[see e.g.][]{Ranalli2003,Gilfanov2004,Laird2005,Mineo2014,Brorby2017}. Observations range typically from 
$\lx/{\rm SFR} \sim 10^{39} - 10^{41} $ \ergs/(\msunyr), with lower (higher) values at high (low) metallicity. As a standard value we adopt $\lx/{\rm SFR}=7.9\times10^{39}$ \ergs/(\msunyr) (hereafter called the ``base model"), and we explore variations of $\pm 1$ dex or more. 
To scale the stellar SED we simply use the standard conversion factor between the UV luminosity
at 1500 \AA\ and SFR, given by \cite{Kennicutt1998}.

The resulting SEDs are then used as input to the Cloudy photoionisation code of \cite{Ferland2017}, version C17.00, to compute a grid of models with varying ionisation parameter with values between $\log U = -1.5$ and $-3.5$. This parameter is proportional to the ratio of ionising photons to total hydrogen density, and the range was chosen in order to span the observed range of the main relative line intensities, as commonly adopted in the literature \citep[e.g.][]{Feltre2017,Ji2020,Ramambason2020}. The models have a closed geometry and are ionisation bounded, i.e. the calculation stops when the hydrogen ionisation front is reached. We adopt solar abundance ratios from \cite{Grevesse2010}, identical stellar and nebular metallicities, and a metallicity of $\oh = 8.1$ for the base model.
The range of parameters explored is described in Table \ref{table_params}.
In this paper the X-ray luminosity, \lx, refers to the energy range of 0.5--8 keV.

   \begin{table}
   \begin{center}
      \caption{Model parameters of the Cloudy grid for the "base model".}
      \small
         \label{table_params}
         \begin{tabular}{ll}
            \hline
            \noalign{\smallskip}
            Parameter      &  Values \\
            \noalign{\smallskip}
            \hline
            \noalign{\smallskip}
            $\log U$ & $-1.5$, $-2.0$, $-2.5$, $-3.0$, $-3.5$  \\
            $\log ((\lx/{\rm SFR})/(\ergs/(\msunyr)))$ & 38.93, 39.90, 40.63, 40.93, 41.18 \\

            \noalign{\smallskip}
            \hline
         \end{tabular}
         \tablefoot{The base model corresponds to a metallicity of 12+log(O/H) = 8.1 and $\lx/{\rm SFR}=7.9\times10^{39}$ \ergs/(\msunyr).}
    \end{center}
   \end{table}

\begin{figure}
   \centering
   \includegraphics[width=\columnwidth]{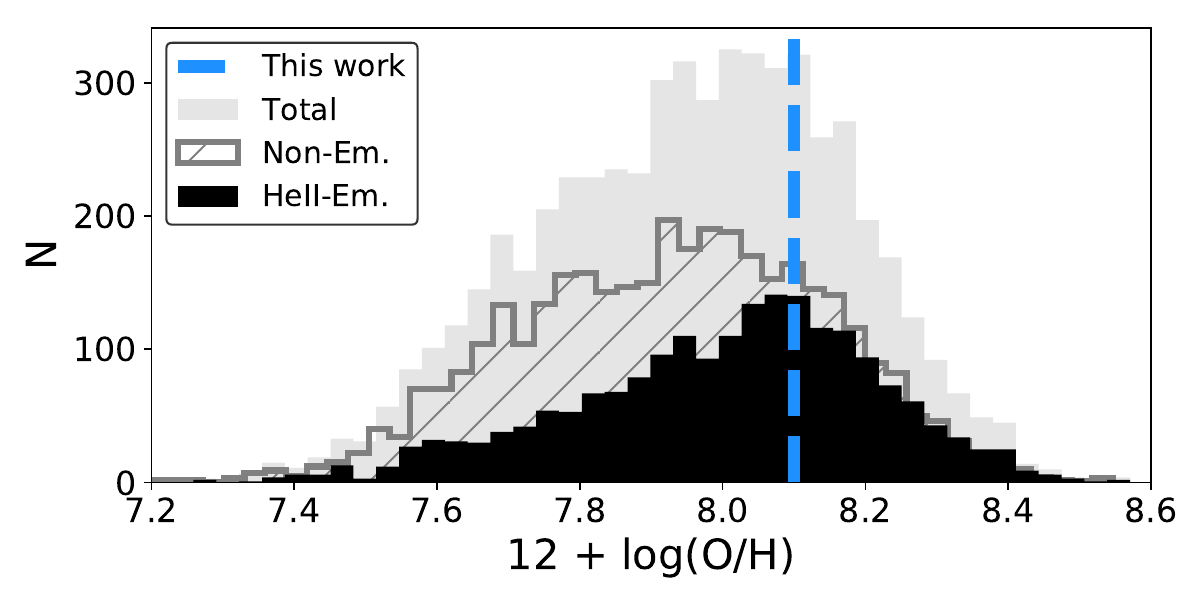}
   \caption{Metallicity histogram for the observational sample. In black we highlight the distribution of the \heii\ emitters. The vertical dashed line shows our selected metallicity, at 12 + log(O/H) = 8.1.}
    \label{OHhist}%
\end{figure}    

\subsection{Observational sample}

To compare our models with observations we use a sample of 5607 star-forming galaxies from the Sloan Digital Sky Survey (SDSS) Data Release 14  \citep{Abolfathi_2018} compiled by Y.\ Izotov and collaborators, analysed in earlier publications \citep{Guseva_2019,Ramambason2020}. The selection criteria used for the extraction of galaxies with active star formation are presented in \cite{Izotov2014}. 
Then we require a detection of the [O~{\sc iii}] $\lambda$4363 line with an accuracy better than $4 \sigma$, allowing thus direct abundance determinations using the $T_e$-method. The overall properties of the parent sample from the Data Release 14 are discussed in \cite{Guseva_2019}. 

In Fig.~\ref{OHhist} we show the metallicity distribution of the observational sample, and the metallicity of our Cloudy models chosen to be representative of the observations.
In this sample, approximately 36 \% show nebular \Heii\ emission, with a mean \heii\ /\hb\ ratio of $I(4686)/I(\hb)= 0.011 ^{+0.007}_{-0.004}$ and a mean metallicity of the \heii\ emitters of $\oh = 8.030 ^{+0.121}_{-0.158}$. The errors indicate the dispersion of the distributions.
In approximately 2/3 of the galaxies HeII emission is not detected, with relative intensities below $I(4686)/I(\hb) \la (2-3) \times 10^{-3}$. 
A smaller, but similarly selected sample of nebular \heii\ emitters has been discussed in \cite{Schaerer2019} and \cite{Schaerer2019b}.

\begin{figure}[tb]
   \centering
   \includegraphics[width=1\columnwidth]{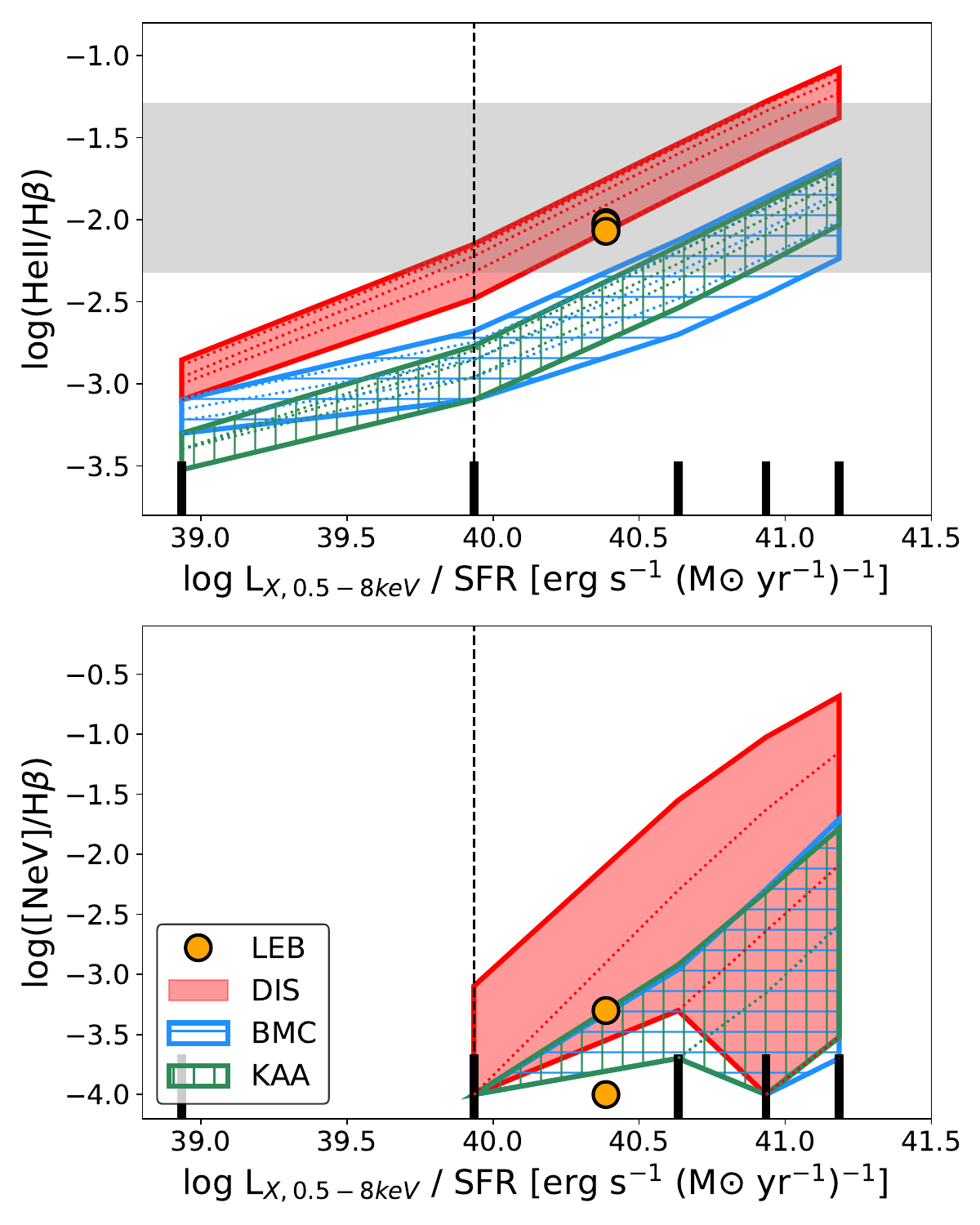}
   \caption{Predicted relative emission line intensity ratios of nebular \Heii\ (top) and \Nev\ (bottom) as a function of \lx/SFR for different SEDs. Each colour represents a different X-ray model. The horizontal gray area shows the typical range of the observed \heii\ line intensity, when detected. The vertical dashed line indicates the \lx/SFR value of our "base model". The dispersion in each SED model is due to changes in the ionisation parameter.
   It can be seen that an increase of the X-ray component produces stronger \heii\ emission, overlapping with the observational area for $\lx/{\rm SFR} \simeq 10^{40}-10^{41}$ \ergs/(\msunyr), depending on the adopted ULX SED. For comparison, the value for the LEB models are overplotted, showing that the LEB SED can reproduce the observed \heii\ strength. In approximately 2/3 of the galaxies HeII emission is not detected, with relative intensities below I(4686)/I(H$\beta$) $\sim (2 - 3) \times 10^{-3}$. The models with $\lx/{\rm SFR}< 10^{40}$ \ergs/(\msunyr) do not produce significant [NeV] emission (i.e. I([NeV])/I(H$\beta$)$ < 10^{-4}$) , and are therefore omitted.}
    \label{fig_senchyna}%
\end{figure}

\section{Results}
\label{sect_results}

We now present the results from our photoionisation models and compare them to observations.

\subsection{Galaxy SEDs including ULXs can produce enough \heii\ emission to explain observations}

In Fig.\ \ref{fig_senchyna} (top panel) we show the predicted \Heii/\hb\ intensity for the Cloudy models using different SEDs, and as a function of \lx/SFR. The range of observed \Heii/\hb\ intensities (when detected) is also shown for comparison. At given \lx/SFR, the highest \Heii/\hb\ intensities are found for the DIS models, followed by BMC, and KAA. This is expected since it reflects the decreasing amount of the \hep\ ionising photons of these SEDs when normalised to the same X-ray luminosity, expressed by $Q_2/\lx$ (see Table \ref{table_params}). The dispersion for a given SED is due to changes in the ionisation parameter, with higher $U$ leading to stronger \heii\ emission.
The LEB models with different log<U> are overplotted, showing that the LEB SED can reproduce the observed \heii\ strength. Moreover, the points overlap showing that the variation of log<U> does not make a significant difference in the predicted \heii\ emission.
In passing we note that the Cloudy predictions for \Heii/\hb\ for SEDs including X-rays  agree or are somewhat lower than the expectations from Case B, as shown in the Appendix.

Finally, \Heii/\hb\ increases with \lx/SFR, since the contribution from X-rays controls the emission of \hep\ ionising photons, which primarily originate from the ULX component (and not from stars).

Compared to observations we find that some of the models predict \heii\ intensities which are strong enough to reach the levels of the observations, i.e.\ galaxies where nebular \heii\ is detected. Overall, with the ULX SEDs examined here, we find that values of $\lx/{\rm SFR} \ga 10^{40}$ \ergs/(\msunyr)\ are required to reach currently detectable levels of \Heii\ emission. There are several ways to explain the absence of \heii\ emission, well below the threshold of the current observations (\Heii/\hb $\la 0.005$). Indeed, even in the presence of ULXs or other X-ray sources, the diversity of their intrinsic properties (SEDs in particular) and variations in the ionisation parameter can cause very weak or no \heii\ emission. More detailed comparisons with observations, involving also other emission lines will be discussed below.

\subsection{\nev\ emission and its relation with \heii}

In addition to \heii, a few blue compact dwarf galaxies show \Nev\ emission \citep{Izotov2004NeV,Thuan2005,Izotov2012}. Given its high ionisation potential of 97.1 eV, this line is often associated to hard non-thermal radiation from AGN, making its detection in star-forming galaxies interesting to explore. The study of \cite{Izotov2012} lists in total eight low-metallicity star-forming galaxies with intensities \Nev/\Heii $\sim 0.2-0.5$, which they propose can be explained by fast radiative shocks contributing $\sim 10$ per cent of the stellar ionising (Lyman continuum) flux. We attempt to explain these observations with our models.      

In Fig.~\ref{fig_senchyna} (bottom panel) we show the predicted \Nev\ line intensity. As expected, for a given \lx/SFR the highest \Nev/\hb\ intensities are again found in the DIS model, followed by BMC and KAA. This is analogous to the case of \heii/\hb, and is explained by the decreasing amount of Ne$^{4+}$ ionising photons in the different SEDs. Compared to \heii/\hb, however, the \nev\ line intensity decreases more rapidly with decreasing \lx/SFR, and a higher \lx/SFR is needed in order to produce \nev\ emission with comparable (and detectable) intensities as \Heii. Finally, the \nev\ emission depends strongly on the ionisation parameter (as expected for forbidden lines), in contrast to the \heii\ recombination lines, and significant emission is only predicted for models with high $U$ ($\log U = -1.5$ to $-2.5$).
The observed intensities, reaching up to \Nev/\hb $=0.0091 \pm 0.0031$ in galaxies with strong \Heii/\hb\ $\sim 0.03$ \citep{Izotov2012} are well reproduced by our models.
The strong dependence of the \nev\ line on $U$ (primarily) and variations in SED shape can also naturally explain why only a (small) fraction of \heii\ emitters also show \nev.

\begin{figure*}
   \centering
   \includegraphics[width=2\columnwidth]{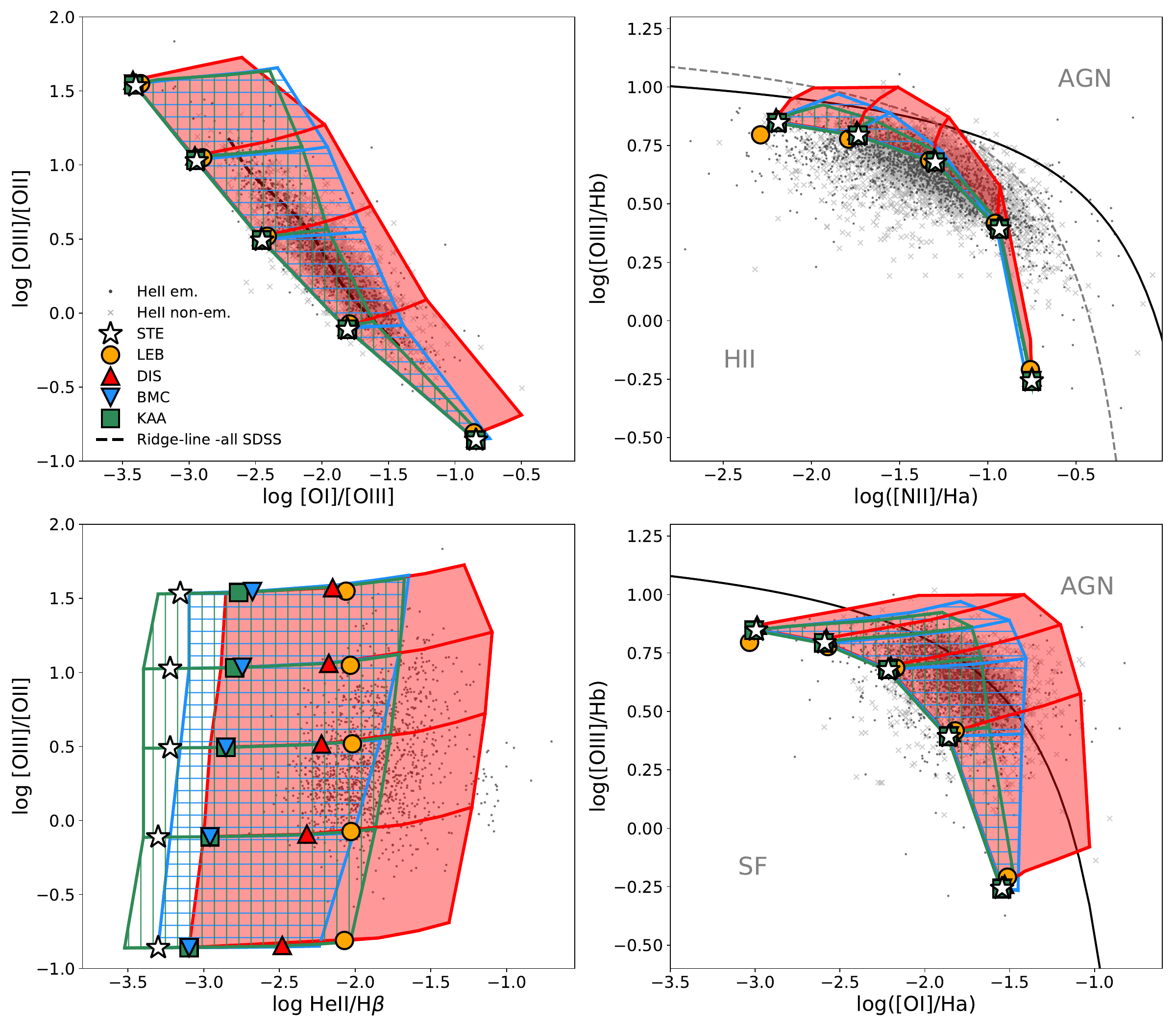}
   \caption{Emission line ratio diagrams for our models ($\log U$ and \lx/SFR) for different SEDs compared to the observational sample. The left column shows high-ionisation emission line diagrams (see \cite{Stasinska2015}), while the right column shows low-ionisation diagrams \citep[BPT diagram; e.g.,][]{Kewley2006,Baldwin1981}. The points within a same model show the outputs obtained for the base models and different ionisation parameters. 
   The shaded/coloured areas show the range of emission line ratios covered by the DIS, BMC and KAA models with varying X-ray luminosities and ionisation parameter.
   The lines show the results for a same ionisation parameter for each model (from top to bottom, $\log U = -1.5, -2.0, -2.5, -3.0, -3.5$) and increasing X-ray emission from left to right.
   For the stellar (STE) and the LEB SED the X-ray luminosity is constant by construction (and $L_X=0$ for STE).
   The observational sample is shown by the grey symbols, where dots show the \heii\ emitters and crosses the sources where \heii\ is not detected. In approximately 2/3 of the galaxies HeII emission is not detected, with relative intensities below I(4686)/I(H$\beta$) $\sim (2 - 3) \times 10^{-3}$.}
           \label{fig_diagrams}%
    \end{figure*}
   
\begin{figure}
   \centering
   \includegraphics[width=\columnwidth]{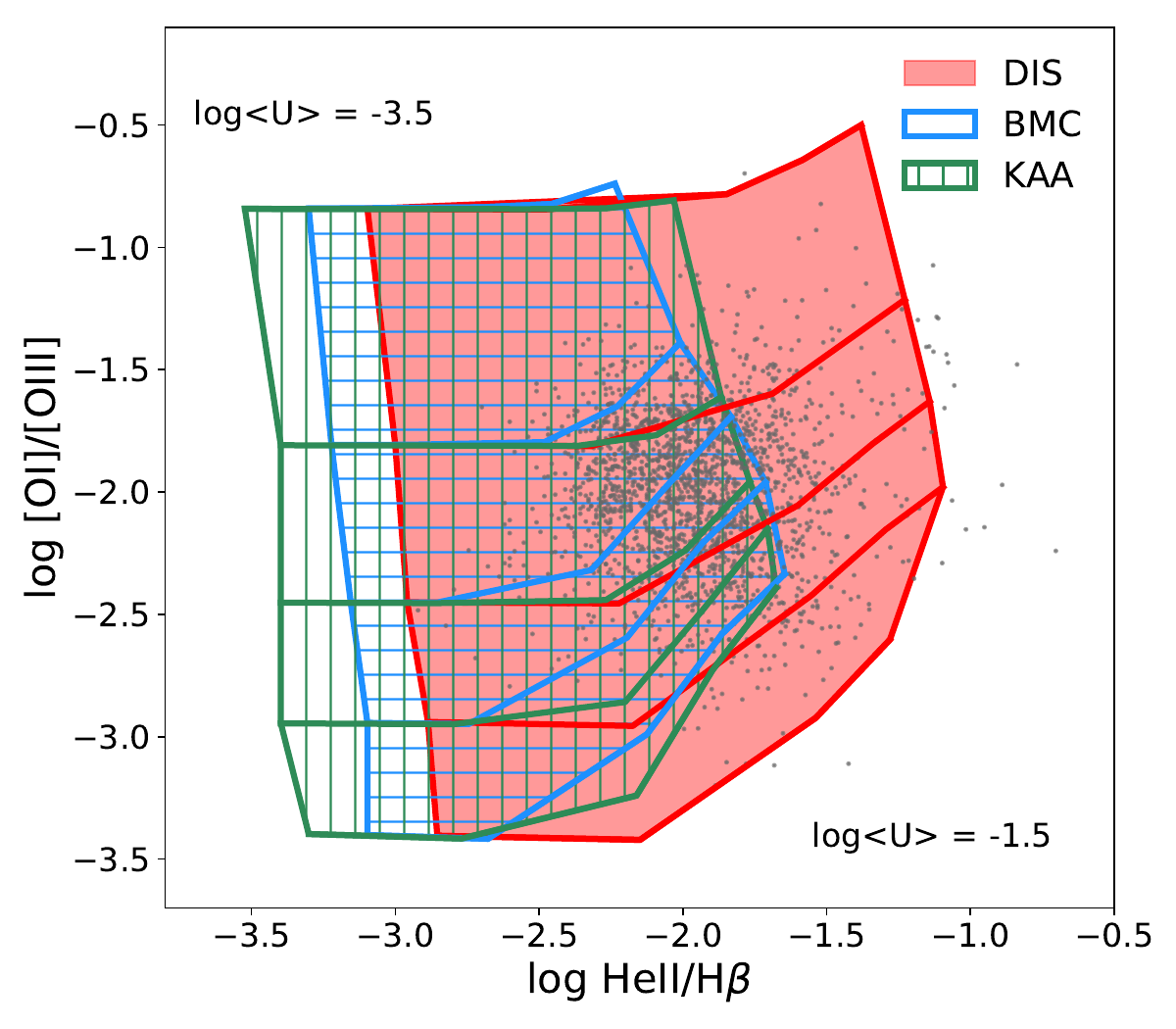}
   \caption{Relation between [O\rm{I}]/[O\rm{III}] and He\rm{II}/H$\beta$ for the DIS, BMC and KAA models (shaded areas) along with the observational sample (dots). From top to bottom we show increasing ionisation parameter (lines). In approximately 2/3 of the galaxies HeII emission is not detected, with relative intensities below I(4686)/I(H$\beta$) $\sim (2 - 3) \times 10^{-3}$.} 
    \label{fig_HeIIOI}%
\end{figure}

\subsection{The impact of ULXs on emission line diagnostics}
To examine the effect of the high energy part of the SED on the optical emission lines we reproduce line ratio diagrams probing three ionisation stages of oxygen, the strength of nebular \heii, and two classical BPT diagrams \citep{Baldwin1981} involving oxygen and nitrogen lines. Figure~\ref{fig_diagrams} shows four different representations of the results: \oiii/\oii\ versus \oi/\oiii\ and \heii/\hb\ on the left panels, and BPT on the right panel. Several of these diagrams were also discussed and compared to the same observational sample in \cite{Stasinska2015} and \cite{Ramambason2020}.
As shown in Fig.\ \ref{fig_diagrams}, and well known, the ionisation parameter $U$ determines basically the observed \oiii/\oii\ ratio, and overall also \oiiil/\hb\ \citep[cf.][]{Kewley2006}. Then, at fixed $U$, the addition of X-rays leads primarily to an increase of \Heii/\hb\ as shown above (Fig.\ \ref{fig_senchyna}), by up to three orders of magnitudes for the model SEDs and maximum $L_X/SFR$ value considered here. 
The low energy tail of the X-ray emission does not only create a central He$^{2+}$ region from which the \heii\ recombination lines are then emitted. The large mean free path of the soft X-ray photons also leads to an extension of the \hii\ region in the outer parts, causing thus stronger emission from low ionisation lines emitted in these regions. This explains the increase of \Oi/\ha\ also shown in this figure. 
As expected, the harder the spectra -- as measured by $Q_2/Q_0$ -- the higher is the maximum intensity of both \Heii/\hb\ and \Oi/\ha. 
Other lines such as \nii\ are not significantly affected by X-rays, because the \nii\ lines are formed in more central regions of the nebula.

The comparison of the models with the observations plotted in Fig.\ \ref{fig_diagrams} shows that the combination of stellar and ULX SEDs allows to cover basically the entire range of emission line ratios, including \heii/\hb\ as already shown in Fig.\ \ref{fig_senchyna}. 
The \Nii/\ha\ BPT diagram is well reproduced, i.e. our models follow the observational sample, which is in particular due to an appropriate N abundance for this sample. 
The purely stellar models and those with low amounts of X-rays do somewhat underpredict the \Oi\ emission, as also seen in other studies \citep{Stasinska2015,Plat_2019,Ramambason2020}. Interestingly, the addition of X-rays allows simultaneously to reach the observed intensities of \Heii\ and a slight ``boost'' of \Oi\ to potentially reconcile these different line ratios in a consistent manner, which is further illustrated in Fig.\ \ref{fig_HeIIOI}. 
However, alternative explanations for the possible discrepancies of \Oi\ and other low ionisation lines, which are not related to high energy emission, exist \citep[see e.g.][]{Stasinska2015,Ramambason2020}.

\section{Discussion}
\label{sect_discuss}

The above results are promising, showing that the presence of ULXs in star-forming galaxies can  -- in some conditions -- explain the observed emission of nebular \heii\ and other lines in low metallicity galaxies. We now discuss these conditions and related assumptions, other studies on the contribution of ULXs/HMXBs, and other approaches to tackle this problem.

This work has shown that X-ray emission can drive the nebular \heii\ emission by using ULX/HMXBs as the sources of the observed high energy radiation. In particular, we found that the model DIS yields the most promising results. In low metallicity galaxies HMXBs are expected to outnumber low mass X-ray binaries by a factor of $\sim 10$ \citep{Douna2015}. Additionally, about 2/3 of the L$_{X, 0.5-8keV}$ per unit SFR is expected to be due to HMXBs \citep{Mineo2014}. If HMXBs are indeed responsible for the nebular \heii\ emission, the excess of these objects at lower metallicity would agree with the observations.

\subsection{The ULX/HMXB -- \heii\ connection in the literature}

Our study follows up on the scenario proposed by \cite{Schaerer2019} and \cite{Schaerer2019b}, who relate the observed anti-correlation between integrated $L_X$/SFR and metallicity of star-forming galaxies \citep{Douna2015,Brorby2016,Lehmer2021} to the same observed trend of nebular \Heii/\hb\ with metallicity. They show that the observations can roughly be reproduced with simple assumptions and an empirical calibration of $Q_2/L_X=(1-3) \times 10^{10}$ photon erg$^{-1}$  based on the ULX observed in the low-metallicity galaxy I Zw 18. 
This work postulates that X-ray bright sources (HMXB, ULX) are the main source of He$^+$ ionising photons in low-metallicity galaxies.

Overall, the present work builds on the same assumptions, using, however, detailed SEDs of ULX to perform quantitative photoionisation models to predict not only the He+ and H recombination lines, but also metallic forbidden lines, which are successfully compared to the observations. In particular, our Cloudy models can reproduce the observed intensities of the high ionisation lines of \heii, \Nev, as well as the lines of \oiii, \oii, and \oi\footnote{Since observations probing O~{\sc iv} lines are quite rare, we are currently not able to examine this ionisation stage as well.}.
Observationally, \cite{Saxena2020AA,Saxena2020MNRAS} have explored the X-ray emission from a sample of 18 \heii\ emitters and a comparison sample of non-\heii\ emitters at redshift $z \sim 3$ by stacking, finding similar ratios of $L_X$/SFR for both samples. From this they conclude that X-ray binaries or weak or obscured AGN are unlikely to be the dominant producers of He$^+$ ionising photons in star-forming galaxies at this redshift. 
For the 18 \heii\ emitters available for this study they obtain a stacked $\log <L_X>/<SFR>=40.01^{+0.26}_{-0.75}$ (in \ergs/(\msunyr) at $z=3.04$, and $39.95^{+0.12}_{-0.17}$ for the non-emitters at $z=3.15$. Both values are compatible with existing measurements at $z \sim 1.5-2.3$ \citep{Fornasini2019}. However, the uncertainties on $L_X/$SFR are quite large, especially for the \heii\ emitters, and the $L_X/$SFR of both samples is not significantly larger than the average value observed in the low-$z$ Universe at solar metallicity $\log L_X/SFR=39.55 \pm 0.15$ \cite[see e.g.][]{Brorby2016}, where nebular \heii\ emission is generally not observed \citep[cf.][]{Guseva2000,Shirazi2012}. Further observational studies are desirable to reach more firm conclusions (cf.\ below).

\cite{Senchyna2020} have done a similar modelling exercise as the present study, but have reached opposite results. They use spectra of HMXB from the multi-colour disc model \citep{Mitsuda1984}, which produces a modified blackbody spectrum representing gas at a range of temperatures in the accretion disc, and different black hole masses. These SEDs are then combined in the same way as in our study to the SED of integrated stellar populations, assuming variable amounts of $L_X/SFR$, and then processed by Cloudy models. From their work, \cite{Senchyna2020} conclude that HMXBs are not efficient enough to produce the needed amount of He$^+$ ionising photons, and hence that the HMXBs are not the dominant source of nebular \heii\ in low-metallicity galaxies. 
The main differences with the present work are the adopted SEDs used to describe the X-rays sources. Clearly, the SEDs assumed by \cite{Senchyna2020} are very hard, and hence predict only small amounts of ionising photons close to the He$^+$ ionisation potential. In other words their input spectra have a significantly lower $Q_2/L_X$ than the ULX SEDs adopted here, which directly translates to lower \Heii/\hb\ ratios for the same amount of X-ray emission (i.e.\ same $L_X/$SFR).
The success or failure of their model and ours therefore largely depends on the assumed X-ray SEDs which are adopted. 
We now discuss this issue and the diversity of SEDs among ULXs and HMXBs.

\subsection{Diversity of SEDs among ULX}
\label{s_diversity}

Early ULX observations were dominated by a single spectral component that could be fitted with accretion disk models \citep{Makishima2000}. However, as the spectral quality of observations improved with \textsl{Chandra} and \textsl{XMM-Newton}, it became evident that this single component model was not physically motivated and that deeper observations required more complex modelling \citep{Kaaret2017}. 

A physically motivated ULX model would have two main components: an accretion disk modelled by radially dependent blackbodies and a Comptonised corona. The relative flux of these components varies, making the spectrum change its hardness \citep[e.g. "soft UL" and "hard UL";][]{Sutton2013a,Sutton2013b}.  The fundamental properties of X-ray binaries are the mass, spin and nature of the compact object and companion star, as well as their orbital separation and eccentricity. The variation of these properties produces different SED shapes.  In addition, as already discussed above, different components may be present (intermediate-mass BHs, irradiated disks, supercritical irradiated disks), and the mass-transfer rate is also expected to affect the intrinsic SEDs.
Many studies suggest that ULXs correspond to a subset of sources with super-Eddington accretion rates seen basically face-on, whereas observations at high inclinations would see the enshrouded accretor as microquasars or super-soft sources \citep[see][and references therein]{Kaaret2017,Urquhart2019}.

Observationally, ULXs show emission across a wide spectral range, including in some cases radio emission and nebulae detected at optical and IR wavelengths, illustrating the many diverse associated processes \citep[e.g.][]{Cseh2014,Liu2015}.
In the case of nebulae, the detected emission lines may provide useful information on the SEDs in the spectral range of interest here (primarily $>54$ eV and up to the X-ray domain) and on contributions from other processes (e.g.\ shocks).
Indeed, as already mentioned in the Introduction, numerous ULX nebulae showing high ionisation lines as \Heii,  \Nev, and [O~{\sc iv}] have been reported \citep[e.g.][]{Pakull2002,Lehmann2005,Abolmasov2007,Abolmasov2008,Kaaret2009,Berghea2010,Grise2011,Maggi2011,Moon2011Large-Highly-Io,Berghea2012,Binder2018,Urquhart2018,Vinokurov2018}, but others do not show such lines \citep[e.g.][]{Pakull2002,Abolmasov2007,DiazTello2017}. Interestingly, some microquasars also show nebulae with \Heii\ emission \citep{Pakull2010}, another property in common with (some) ULX. Statistical results regarding the frequency and strength of these important lines have, to the best of our knowledge, not been reported. In any case, the present results point towards a diversity of the SED in the UV-to-X-ray domain.

Relatively few studies so far have examined ULX nebulae in classical BPT diagrams or used similar diagnostics to determine the dominant source of the optical emission \citep[see e.g.][]{Lehmann2005,Abolmasov2008}. Recent examples of BPT diagrams are shown e.g.\ by \citep{DiazTello2017}, who find a ULX nebula located in the borderline between star forming and AGN regions. Another nebula is found in the LINER domain \citep{Lopez2019}. The emission line ratios of the high-ionisation nebula Holmberg II X-1 are consistent with \hii\ regions \citep{Lehmann2005}. According to \cite{Kaaret2017}, most nebulae appear to be powered by shocks between outflows and the surrounding medium, while few are powered by photoionisation. Again, both X-ray/photo-ionised regions and shock-ionised nebulae seem to exist, but statistics and a more profound understanding of these nebulae is lacking.

In short, the variety and range of properties exhibited by ULXs make the modelling of ULXs a challenging task. 

\subsection{Other sources of nebular \heii\ emission}
\label{sect_other}
Several other processes and sources to explain nebular \heii\ emission in low-metallicity galaxies have been suggested and explored in the literature.
These include for example shocks in the ISM and the existence of very hot stars, in addition to other ``unconvential'' suggestions. 

\cite{Thuan2005}, \cite{Izotov2012}, and others have proposed that fast radiative shocks can explain both the observed \Heii\ and \Nev\ intensities in some low-metallicity galaxies. Using the shock models of \cite{Allen2008} for SMC and LMC metallicities, \cite{Izotov2012} show that the main intensities can be reproduced with shocks contributing $\sim 10$ \% of the total ionising flux, the rest being due to stars. Observed broad components in strong emission lines such as \ha\ could be a direct observational signature indicating such shocks. 
The studies of \cite{Plat_2019} and others reach similar conditions, arguing for shock or an AGN contribution to explain nebular \heii.
The principle difficulties of this scenario are probably to predict the required shock contribution in a quantitative manner, and to explain the observed increase of \heii/\hb\ with decreasing metallicity. To the best of our knowledge this has not yet been addressed in the literature.

Basically since the discovery of frequent nebular \heii\ emission in low-metallicity galaxies, it has been shown that state-of-the-art evolutionary synthesis models combining the latest knowledge of stellar evolution and stellar atmospheres fall short of producing the high energy ($> 54$ eV) photons needed to reproduce this high ionisation line (see references in Sect.\ 1). Despite several updates, improvements, and the inclusion of stars resulting from stellar interactions in binary systems, this conclusion remains valid so far, as shown by several recent studies examining both the UV and/or optical \heii\ recombination lines \citep[e.g.][]{Berg2018,Stanway2019,Nanayakkara2019,Saxena2020AA,Bian2020}.

Despite this, several authors suggest that binary evolution, fast rotation or other processes could lead to very hot stars, which should boost and maybe provide sufficient He$^+$ ionising photons \citep[e.g.][]{Gotberg2019,Bian2020}. However, no quantitative stellar population model solving this observational problem has been presented so far.
Two recent studies propose that nebular \heii\ emission with the observed intensities could even be achieved with conventional stellar populations, i.e.\ without the need to invoke other sources of higher energy photons.
For example, \cite{Barrow2019} performed cosmological zoom-in simulations of high-redshift galaxies including SEDs from the Flexible Stellar Population Synthesis code \citep[FSPS][]{Conroy2010} and following the evolution on very short timescales. In these simulations they identify several galaxies with \Heii/\hb $\sim 0.001-0.03$, which are found during specific, relatively short time intervals. The study of \cite{Barrow2019} argues that gaps of $\geq 15$ Myr in star formation could explain the high intensities of \heii\ without the need for other ionising sources.

However, in the detailed simulations which reach metallicities of \oh $\la 8.0$, comparable to those observed in the low-$z$ galaxy sample, their models predict lower intensities (\Heii/H$\beta$ $\la 0.007$) than the observations. 
Furthermore, the predicted \heii\ phases occur only in sources with low \hb\ equivalent widths (EW(\hb) $\la 10$ \AA), in stark contrast with observations \citep[see e.g.\ the compilation in][]{Schaerer2019,Schaerer2019b}. 
Finally, the simulations predict only very few sources and short phases with nebular \heii, again in contrast with the observations where nebular \heii\ is detected in approximately $\sim 2/3$ of the star-forming galaxy samples at metallicities of $\oh \la 8.3$.
From this we conclude that their model is not comparable to the observed low-$z$ galaxy sample or that additional high-energy sources are indeed needed.

In a recent work, \cite{PerezMontero2020} have studied the behaviour of the nebular He~{\sc i} and \heii\ lines in low-$z$ galaxies
and suggested that the observed intensities of \heii\ could be explained by density-bounded \hii\ regions powered by ``normal'' stellar SEDs. To achieve this with normal SEDs the relative \Heii/\hb\ predicted for Case B (ionisation bounded nebulae) must be significantly ``boosted''. This can in principle be achieved by reducing a large fraction of the \hb\ emission which is emitted quite uniformly across the entire \hii\ region, whereas the central \heii-emitting region remains unchanged. However, to reach the observed \heii\ intensities, a very high mean escape fraction $f_{\rm esc}=0.74$ (or a mean absorption fraction 26 \%) of the H ionising photons must be invoked \citep{PerezMontero2020}, which is not compatible with the observations.
Indeed, overall Lyman continuum escape appears to be relatively rare and certainly well below this value at any redshift $z \la 3$ where it has currently been measured \citep[see e.g.][]{Siana2015,Grazian2017,Steidel2018,Alavi2020}. 
Also, at $z\sim 0.3-0.4$ where recent studies with HST have detected some Lyman continuum emitters, the measured escape fractions range from 
$\sim 2-70$ \% \citep[see e.g.][and references therein]{Izotov2021}, with a mean escape fraction well below value invoked by \cite{PerezMontero2020}.
Furthermore, the known low-$z$ leakers do not show different/enhanced \heii/\hb\ intensities \citep{Guseva2000}.
Finally, with such a large mean escape fraction the observed \hb\ EW distribution of the \heii\ emitters would be significantly altered
(i.e.\ shifted to lower EW$(\hb)$ by a factor $\sim 4$) and many \heii\ emitters showing high EW(\hb) 
\citep[cf.][]{Shirazi2012,Schaerer2019b} could hence not be explained by the same stellar population model which predict a maximum
EW(\hb) which is quite close to the maximum observed value.

\subsection{Future avenues}

From the above discussions we conclude -- based on our current knowledge of stars and stellar populations --  that shocks, ULX/HMBX, a low-luminosity AGN contribution, or other sources of high energy photons are needed to solve the nebular \heii\ problem.
None of the existing scenarios provide so far fully satisfactory answers and/or quantitative results.
Clearly, more research is needed to unveil the mystery of \heii\ nebular emission. 
 
For example, theoretical and observational studies to better understand the SEDs of ULX and HMXBs and their diversity might be of interest. Similarly, a better understanding of ULX nebulae, their frequency, observed spectra etc.\ would be useful.
More generally, investigations on the impact of X-ray sources (both discrete and diffuse) on the ISM and observed spectra of star-forming galaxies may also shed new light on the \heii\ problem and other, possibly related issues. Along these lines, several recent studies have e.g.\ 
investigated the link between X-rays and Ly$\alpha$ emitters, and possible connections between X-ray sources and Lyman continuum escape
\citep[see e.g.][]{Sobral2015,Svoboda2019,Dittenber2020}.
More detailed investigations of shocks and their contribution of high ionisation lines, both on small scales (resolved) as well as in integrated galaxy spectra, would also seem of considerable interest. 

The \textsl{James Webb Space Telescope} and other large telescopes will certainly continue to detect nebular \heii\ emission in progressively more distant galaxies and provide invaluable new information on the stars, the ISM, and other components of galaxies in the early Universe. 
A proper solution of the long-standing nebular \heii\ problem, which may affect other emission line diagnostics to a currently unknown extent,
appears as a fundamental goal to confidently exploit these powerful new observations.

\section{Conclusions}
\label{sect_conclude}

We have quantitatively examined the impact of luminous X-ray sources on the optical emission lines of star-forming galaxies and nebular \heii\ in particular. To do so we have combined three empirically-constrained SEDs of ULXs with known nebulae from \cite{Berghea2012} and \cite{Kaaret2009} with the SED of stellar populations, as predicted from BPASS models \citep{Eldridge2017}. The X-ray luminosity of the source normalised to the SFR of the galaxy, $L_X/$SFR, is considered a free parameter, which is used to explore the effect of the X-ray sources. We varied $L_X/$SFR from $10^{39}$ to $\sim 10^{41}$\ergs/(\msunyr), covering thus the range observed in star-forming galaxies \citep[e.g.][]{Mineo2014,Brorby2017}.
Using the combined SEDs we have then computed Cloudy photoionisation models for conditions representative of low-metallicity galaxies from the sample of \cite{Guseva_2019}.
Our main findings can be summarised as follows:

\begin{itemize}
\item The addition of X-ray emission from a ULX to a purely stellar SED hardens the spectrum of the galaxy, allowing a nebular \heii\ emission with intensities up to \Heii/\hb $\sim 0.04$, comparable to the observations, with X-ray luminosities also compatible with observations.
    Increasing the X-ray contribution directly increases the output \heii\ nebular emission.
\item \nev\ emission, which requires ionising photons with energies above 97.1 eV and is observed together with \heii\ in a small number of blue compact dwarf galaxies, can be explained by the same models. The strong dependence of \nev\ on the ionisation parameter can naturally explain why this line is only detected in a small fraction of \heii\ emitters.
\item The contribution of ULXs to the galaxy SED affects the classical optical diagnostic diagrams. As expected, the high energy radiation boosts both high and low ionisation lines. We show, e.g., that the BPT diagram involving \Oi/\ha\ is more significantly altered than \Nii/\ha.
\item  We show that the main optical emission line ratios, involving three ionisation stages of Oxygen (namely \Oiii, \Oii, and \Oi), \Heii, \Nii, plus \ha\ and \hb, can consistently be reproduced by our galaxy SEDs including an ULX component.
\item In particular, we have shown that the most efficient model is one in which an irradiated disc X-ray model is added, which can fit some ULXs both optical and X-ray wavelengths (DIS).
The base spectra for the DIS model can achieve the observed percentage, while the BMC and KAA models are below expected.
\item The impact of ULXs on the UV-optical emission lines of low-metallicity star-forming galaxies depends strongly on the shape of the SEDs of these X-ray sources, which are poorly known, most likely variable, and thus difficult to describe in an average sense.
\end{itemize}

\begin{acknowledgements}
We thank Ciprian Berghea, Philipp Kaaret, and Vianney Lebouteiller for communicating SEDs from their work.
We also thank various colleagues, including Yuri Izotov, Aayush Saxena, Laura Pentericci, Tassos Fragos, Andr\'es G\'urpide, Lida Oskinova, Nathalie Webb, Devina Misra, and John Chisholm for stimulating discussions on the nebular \heii\ problem and ULXs during recent years. C.S and A.V. acknowledge support from SNF Professorship PP00P2\_176808. A.V. is supported by the ERC starting grant ERC-757258-TRIPLE.
\end{acknowledgements}

\bibliography{bib}

\begin{thebibliography}{120}
\expandafter\ifx\csname natexlab\endcsname\relax\def\natexlab#1{#1}\fi

\bibitem[{{Abolfathi} {et~al.}(2018){Abolfathi}, {Aguado}, {Aguilar}, {Allende
  Prieto}, {Almeida}, {Ananna}, {Anders}, {Anderson}, {Andrews}, {Anguiano}, \&
  et~al.}]{Abolfathi_2018}
{Abolfathi}, B., {Aguado}, D.~S., {Aguilar}, G., {et~al.} 2018, ApJS, 235, 42

\bibitem[{{Abolmasov} {et~al.}(2007){Abolmasov}, {Fabrika}, {Sholukhova}, \&
  {Afanasiev}}]{Abolmasov2007}
{Abolmasov}, P., {Fabrika}, S., {Sholukhova}, O., \& {Afanasiev}, V. 2007,
  Astrophysical Bulletin, 62, 36

\bibitem[{{Abolmasov} {et~al.}(2008){Abolmasov}, {Fabrika}, {Sholukhova}, \&
  {Kotani}}]{Abolmasov2008}
{Abolmasov}, P., {Fabrika}, S., {Sholukhova}, O., \& {Kotani}, T. 2008, arXiv
  e-prints, arXiv:0809.0409

\bibitem[{{Alavi} {et~al.}(2020){Alavi}, {Colbert}, {Teplitz}, {Siana},
  {Scarlata}, {Rutkowski}, {Mehta}, {Henry}, {Dai}, {Haardt}, \&
  {Bagley}}]{Alavi2020}
{Alavi}, A., {Colbert}, J., {Teplitz}, H.~I., {et~al.} 2020, \apj, 904, 59

\bibitem[{{Allen} {et~al.}(2008){Allen}, {Groves}, {Dopita}, {Sutherland}, \&
  {Kewley}}]{Allen2008}
{Allen}, M.~G., {Groves}, B.~A., {Dopita}, M.~A., {Sutherland}, R.~S., \&
  {Kewley}, L.~J. 2008, \apjs, 178, 20

\bibitem[{{Arnaud}(1996)}]{Arnaud1996}
{Arnaud}, K.~A. 1996, in Astronomical Society of the Pacific Conference Series,
  Vol. 101, Astronomical Data Analysis Software and Systems V, ed. G.~H.
  {Jacoby} \& J.~{Barnes}, 17

\bibitem[{{Baldwin} {et~al.}(1981){Baldwin}, {Phillips}, \&
  {Terlevich}}]{Baldwin1981}
{Baldwin}, J.~A., {Phillips}, M.~M., \& {Terlevich}, R. 1981, \pasp, 93, 5

\bibitem[{{Barrow}(2019)}]{Barrow2019}
{Barrow}, K. S.~S. 2019, \mnras, 2947

\bibitem[{{Berg} {et~al.}(2019){Berg}, {Chisholm}, {Erb}, {Pogge}, {Henry}, \&
  {Olivier}}]{Berg2019}
{Berg}, D.~A., {Chisholm}, J., {Erb}, D.~K., {et~al.} 2019, \apjl, 878, L3

\bibitem[{{Berg} {et~al.}(2021){Berg}, {Chisholm}, {Erb}, {Skillman}, {Pogge},
  \& {Olivier}}]{Berg2021}
{Berg}, D.~A., {Chisholm}, J., {Erb}, D.~K., {et~al.} 2021, arXiv e-prints,
  arXiv:2105.12765

\bibitem[{{Berg} {et~al.}(2018){Berg}, {Erb}, {Auger}, {Pettini}, \&
  {Brammer}}]{Berg2018}
{Berg}, D.~A., {Erb}, D.~K., {Auger}, M.~W., {Pettini}, M., \& {Brammer}, G.~B.
  2018, \apj, 859, 164

\bibitem[{{Berghea} \& {Dudik}(2012)}]{Berghea2012}
{Berghea}, C.~T. \& {Dudik}, R.~P. 2012, \apj, 751, 104

\bibitem[{{Berghea} {et~al.}(2010){Berghea}, {Dudik}, {Weaver}, \&
  {Kallman}}]{Berghea2010}
{Berghea}, C.~T., {Dudik}, R.~P., {Weaver}, K.~A., \& {Kallman}, T.~R. 2010,
  \apj, 708, 364

\bibitem[{{Bian} {et~al.}(2020){Bian}, {Kewley}, {Groves}, \&
  {Dopita}}]{Bian2020}
{Bian}, F., {Kewley}, L.~J., {Groves}, B., \& {Dopita}, M.~A. 2020, \mnras,
  493, 580

\bibitem[{Binder {et~al.}(2018)Binder, Levesque, \&
  Dorn-Wallenstein}]{Binder2018}
Binder, B., Levesque, E.~M., \& Dorn-Wallenstein, T. 2018, The Astrophysical
  Journal, 863, 141

\bibitem[{{Brorby} \& {Kaaret}(2017)}]{Brorby2017}
{Brorby}, M. \& {Kaaret}, P. 2017, \mnras, 470, 606

\bibitem[{{Brorby} {et~al.}(2016){Brorby}, {Kaaret}, {Prestwich}, \&
  {Mirabel}}]{Brorby2016}
{Brorby}, M., {Kaaret}, P., {Prestwich}, A., \& {Mirabel}, I.~F. 2016, \mnras,
  457, 4081

\bibitem[{{Cassata} {et~al.}(2013){Cassata}, {Le F{\`e}vre}, {Charlot},
  {Contini}, {Cucciati}, {Garilli}, {Zamorani}, {Adami}, {Bardelli}, {Le Brun},
  {Lemaux}, {Maccagni}, {Pollo}, {Pozzetti}, {Tresse}, {Vergani}, {Zanichelli},
  \& {Zucca}}]{Cassata2013}
{Cassata}, P., {Le F{\`e}vre}, O., {Charlot}, S., {et~al.} 2013, \aap, 556, A68

\bibitem[{{Conroy} \& {Gunn}(2010)}]{Conroy2010}
{Conroy}, C. \& {Gunn}, J.~E. 2010, {FSPS: Flexible Stellar Population
  Synthesis}

\bibitem[{Cseh {et~al.}(2012)Cseh, Corbel, Kaaret, Lang, Gris{\'{e}}, Paragi,
  Tzioumis, Tudose, \& Feng}]{Cseh2012}
Cseh, D., Corbel, S., Kaaret, P., {et~al.} 2012, The Astrophysical Journal,
  749, 17

\bibitem[{{Cseh} {et~al.}(2014){Cseh}, {Kaaret}, {Corbel}, {Grise}, {Lang},
  {Kording}, {Falcke}, {Jonker}, {Miller-Jones}, {Farrell}, {Yang}, {Paragi},
  \& {Frey}}]{Cseh2014}
{Cseh}, D., {Kaaret}, P., {Corbel}, S., {et~al.} 2014, \mnras, 439, L1

\bibitem[{{D{\'\i}az Tello} {et~al.}(2017){D{\'\i}az Tello}, {Miyaji},
  {Ishigaki}, {Krumpe}, {Ueda}, {Brunner}, {Goto}, {Hanami}, \&
  {Toba}}]{DiazTello2017}
{D{\'\i}az Tello}, J., {Miyaji}, T., {Ishigaki}, T., {et~al.} 2017, \aap, 604,
  A14

\bibitem[{{Dittenber} {et~al.}(2020){Dittenber}, {Oey}, {Hodges-Kluck},
  {Gallo}, {Hayes}, {{\"O}stlin}, \& {Melinder}}]{Dittenber2020}
{Dittenber}, B., {Oey}, M.~S., {Hodges-Kluck}, E., {et~al.} 2020, \apjl, 890,
  L12

\bibitem[{{Douna} {et~al.}(2015){Douna}, {Pellizza}, {Mirabel}, \&
  {Pedrosa}}]{Douna2015}
{Douna}, V.~M., {Pellizza}, L.~J., {Mirabel}, I.~F., \& {Pedrosa}, S.~E. 2015,
  \aap, 579, A44

\bibitem[{{Eldridge} {et~al.}(2017){Eldridge}, {Stanway}, {Xiao}, {McClelland
  }, {Taylor}, {Ng}, {Greis}, \& {Bray}}]{Eldridge2017}
{Eldridge}, J.~J., {Stanway}, E.~R., {Xiao}, L., {et~al.} 2017, \pasa, 34, e058

\bibitem[{{Erb} {et~al.}(2010){Erb}, {Pettini}, {Shapley}, {Steidel}, {Law}, \&
  {Reddy}}]{Erb2010}
{Erb}, D.~K., {Pettini}, M., {Shapley}, A.~E., {et~al.} 2010, \apj, 719, 1168

\bibitem[{{Feltre} {et~al.}(2017){Feltre}, {Charlot}, {Mignoli}, {Bongiorno},
  {Calura}, {Chevallard}, {Curtis-Lake}, {Gilli}, \& {Plat}}]{Feltre2017}
{Feltre}, A., {Charlot}, S., {Mignoli}, M., {et~al.} 2017, Frontiers in
  Astronomy and Space Sciences, 4, 32

\bibitem[{{Ferland} {et~al.}(2017){Ferland}, {Chatzikos}, {Guzm{\'a}n},
  {Lykins}, {van Hoof}, {Williams}, {Abel}, {Badnell}, {Keenan}, {Porter}, \&
  {Stancil}}]{Ferland2017}
{Ferland}, G.~J., {Chatzikos}, M., {Guzm{\'a}n}, F., {et~al.} 2017, \rmxaa, 53,
  385

\bibitem[{{Fern{\'a}ndez-Ontiveros} {et~al.}(2016){Fern{\'a}ndez-Ontiveros},
  {Spinoglio}, {Pereira-Santaella}, {Malkan}, {Andreani}, \&
  {Dasyra}}]{Fernandez2016}
{Fern{\'a}ndez-Ontiveros}, J.~A., {Spinoglio}, L., {Pereira-Santaella}, M.,
  {et~al.} 2016, \apjs, 226, 19

\bibitem[{{Fornasini} {et~al.}(2019){Fornasini}, {Kriek}, {Sanders}, {Shivaei},
  {Civano}, {Reddy}, {Shapley}, {Coil}, {Mobasher}, {Siana}, {Aird}, {Azadi},
  {Freeman}, {Leung}, {Price}, {Fetherolf}, {Zick}, \& {Barro}}]{Fornasini2019}
{Fornasini}, F.~M., {Kriek}, M., {Sanders}, R.~L., {et~al.} 2019, \apj, 885, 65

\bibitem[{{Fragos} {et~al.}(2013){Fragos}, {Lehmer}, {Tremmel}, {Tzanavaris},
  {Basu-Zych}, {Belczynski}, {Hornschemeier}, {Jenkins}, {Kalogera}, {Ptak}, \&
  {Zezas}}]{Fragos2013}
{Fragos}, T., {Lehmer}, B., {Tremmel}, M., {et~al.} 2013, \apj, 764, 41

\bibitem[{{Gierli{\'n}ski} {et~al.}(2009){Gierli{\'n}ski}, {Done}, \&
  {Page}}]{Gierlinski2009}
{Gierli{\'n}ski}, M., {Done}, C., \& {Page}, K. 2009, \mnras, 392, 1106

\bibitem[{{Gilfanov} {et~al.}(2004){Gilfanov}, {Grimm}, \&
  {Sunyaev}}]{Gilfanov2004}
{Gilfanov}, M., {Grimm}, H.~J., \& {Sunyaev}, R. 2004, \mnras, 347, L57

\bibitem[{{Gladstone} {et~al.}(2009){Gladstone}, {Roberts}, \&
  {Done}}]{Gladstone2009}
{Gladstone}, J.~C., {Roberts}, T.~P., \& {Done}, C. 2009, \mnras, 397, 1836

\bibitem[{{Gliozzi} {et~al.}(2011){Gliozzi}, {Titarchuk}, {Satyapal}, {Price},
  \& {Jang}}]{Gliozzi2011}
{Gliozzi}, M., {Titarchuk}, L., {Satyapal}, S., {Price}, D., \& {Jang}, I.
  2011, \apj, 735, 16

\bibitem[{{Goad} {et~al.}(2006){Goad}, {Roberts}, {Reeves}, \&
  {Uttley}}]{Goad2006}
{Goad}, M.~R., {Roberts}, T.~P., {Reeves}, J.~N., \& {Uttley}, P. 2006, \mnras,
  365, 191

\bibitem[{{G{\"o}tberg} {et~al.}(2019){G{\"o}tberg}, {de Mink}, {Groh},
  {Leitherer}, \& {Norman}}]{Gotberg2019}
{G{\"o}tberg}, Y., {de Mink}, S.~E., {Groh}, J.~H., {Leitherer}, C., \&
  {Norman}, C. 2019, \aap, 629, A134

\bibitem[{{Grazian} {et~al.}(2017){Grazian}, {Giallongo}, {Paris}, {Boutsia},
  {Dickinson}, {Santini}, {Windhorst}, {Jansen}, {Cohen}, {Ashcraft},
  {Scarlata}, {Rutkowski}, {Vanzella}, {Cusano}, {Cristiani}, {Giavalisco},
  {Ferguson}, {Koekemoer}, {Grogin}, {Castellano}, {Fiore}, {Fontana},
  {Marchi}, {Pedichini}, {Pentericci}, {Amor{\'\i}n}, {Barro}, {Bonchi},
  {Bongiorno}, {Faber}, {Fumana}, {Galametz}, {Guaita}, {Kocevski}, {Merlin},
  {Nonino}, {O'Connell}, {Pilo}, {Ryan}, {Sani}, {Speziali}, {Testa}, {Weiner},
  \& {Yan}}]{Grazian2017}
{Grazian}, A., {Giallongo}, E., {Paris}, D., {et~al.} 2017, \aap, 602, A18

\bibitem[{{Grevesse} {et~al.}(2010){Grevesse}, {Asplund}, {Sauval}, \&
  {Scott}}]{Grevesse2010}
{Grevesse}, N., {Asplund}, M., {Sauval}, A.~J., \& {Scott}, P. 2010, \apss,
  328, 179

\bibitem[{Gris{\'{e}} {et~al.}(2012)Gris{\'{e}}, Kaaret, Corbel, Feng, Cseh, \&
  Tao}]{Grise2012}
Gris{\'{e}}, F., Kaaret, P., Corbel, S., {et~al.} 2012, The Astrophysical
  Journal, 745, 123

\bibitem[{{Gris{\'e}} {et~al.}(2011){Gris{\'e}}, {Kaaret}, {Pakull}, \&
  {Motch}}]{Grise2011}
{Gris{\'e}}, F., {Kaaret}, P., {Pakull}, M.~W., \& {Motch}, C. 2011, \apj, 734,
  23

\bibitem[{{G{\'u}rpide} {et~al.}(2021{\natexlab{a}}){G{\'u}rpide}, {Godet},
  {Koliopanos}, {Webb}, \& {Olive}}]{Gurpide2021a}
{G{\'u}rpide}, A., {Godet}, O., {Koliopanos}, F., {Webb}, N., \& {Olive}, J.~F.
  2021{\natexlab{a}}, \aap, 649, A104

\bibitem[{{G{\'u}rpide} {et~al.}(2021{\natexlab{b}}){G{\'u}rpide}, {Godet},
  {Vasilopoulos}, {Webb}, \& {Olive}}]{Gurpide2021b}
{G{\'u}rpide}, A., {Godet}, O., {Vasilopoulos}, G., {Webb}, N.~A., \& {Olive},
  J.~F. 2021{\natexlab{b}}, arXiv e-prints, arXiv:2106.05708

\bibitem[{{Guseva} {et~al.}(2019){Guseva}, {Izotov}, {Fricke}, \&
  {Henkel}}]{Guseva_2019}
{Guseva}, N.~G., {Izotov}, Y.~I., {Fricke}, K.~J., \& {Henkel}, C. 2019, \aap,
  624, A21

\bibitem[{{Guseva} {et~al.}(2000){Guseva}, {Izotov}, \& {Thuan}}]{Guseva2000}
{Guseva}, N.~G., {Izotov}, Y.~I., \& {Thuan}, T.~X. 2000, \apj, 531, 776

\bibitem[{{Izotov} {et~al.}(2014){Izotov}, {Guseva}, {Fricke}, \&
  {Henkel}}]{Izotov2014}
{Izotov}, Y.~I., {Guseva}, N.~G., {Fricke}, K.~J., \& {Henkel}, C. 2014, \aap,
  561, A33

\bibitem[{{Izotov} {et~al.}(2004){Izotov}, {Noeske}, {Guseva}, {Papaderos},
  {Thuan}, \& {Fricke}}]{Izotov2004NeV}
{Izotov}, Y.~I., {Noeske}, K.~G., {Guseva}, N.~G., {et~al.} 2004, \aap, 415,
  L27

\bibitem[{{Izotov} {et~al.}(2012){Izotov}, {Thuan}, \& {Privon}}]{Izotov2012}
{Izotov}, Y.~I., {Thuan}, T.~X., \& {Privon}, G. 2012, \mnras, 427, 1229

\bibitem[{{Izotov} {et~al.}(2021){Izotov}, {Worseck}, {Schaerer}, {Guseva},
  {Chisholm}, {Thuan}, {Fricke}, \& {Verhamme}}]{Izotov2021}
{Izotov}, Y.~I., {Worseck}, G., {Schaerer}, D., {et~al.} 2021, \mnras, 503,
  1734

\bibitem[{{Ji} {et~al.}(2020){Ji}, {Yan}, {Riffel}, {Drory}, \&
  {Zhang}}]{Ji2020}
{Ji}, X., {Yan}, R., {Riffel}, R., {Drory}, N., \& {Zhang}, K. 2020, \mnras,
  496, 1262

\bibitem[{{Kaaret} \& {Corbel}(2009)}]{Kaaret2009}
{Kaaret}, P. \& {Corbel}, S. 2009, \apj, 697, 950

\bibitem[{{Kaaret} {et~al.}(2017){Kaaret}, {Feng}, \& {Roberts}}]{Kaaret2017}
{Kaaret}, P., {Feng}, H., \& {Roberts}, T.~P. 2017, \araa, 55, 303

\bibitem[{{Kehrig} {et~al.}(2021){Kehrig}, {Guerrero}, {V{\'\i}lchez}, \&
  {Ramos-Larios}}]{Kehrig2021}
{Kehrig}, C., {Guerrero}, M.~A., {V{\'\i}lchez}, J.~M., \& {Ramos-Larios}, G.
  2021, \apjl, 908, L54

\bibitem[{{Kennicutt}(1998)}]{Kennicutt1998}
{Kennicutt}, Robert~C., J. 1998, \araa, 36, 189

\bibitem[{{Kewley} {et~al.}(2006){Kewley}, {Groves}, {Kauffmann}, \&
  {Heckman}}]{Kewley2006}
{Kewley}, L.~J., {Groves}, B., {Kauffmann}, G., \& {Heckman}, T. 2006, \mnras,
  372, 961

\bibitem[{{Kovlakas} {et~al.}(2020){Kovlakas}, {Zezas}, {Andrews}, {Basu-Zych},
  {Fragos}, {Hornschemeier}, {Lehmer}, \& {Ptak}}]{Kovlakas2020}
{Kovlakas}, K., {Zezas}, A., {Andrews}, J.~J., {et~al.} 2020, \mnras, 498, 4790

\bibitem[{{Laird} {et~al.}(2005){Laird}, {Nandra}, {Adelberger}, {Steidel}, \&
  {Reddy}}]{Laird2005}
{Laird}, E.~S., {Nandra}, K., {Adelberger}, K.~L., {Steidel}, C.~C., \&
  {Reddy}, N.~A. 2005, \mnras, 359, 47

\bibitem[{{Le F{\`e}vre} {et~al.}(2019){Le F{\`e}vre}, {Lemaux}, {Nakajima},
  {Schaerer}, {Talia}, {Zamorani}, {Cassata}, {Garilli}, {Maccagni},
  {Pentericci}, {Tasca}, {Zucca}, {Amorin}, {Bardelli}, {Cimatti},
  {Giavalisco}, {Guaita}, {Hathi}, {Marchi}, {Vanzella}, {Vergani}, \&
  {Dunlop}}]{LeFevre2019}
{Le F{\`e}vre}, O., {Lemaux}, B.~C., {Nakajima}, K., {et~al.} 2019, \aap, 625,
  A51

\bibitem[{{Lebouteiller} {et~al.}(2017){Lebouteiller}, {P{\'e}quignot},
  {Cormier}, {Madden}, {Pakull}, {Kunth}, {Galliano}, {Chevance}, {Heap},
  {Lee}, \& {Polles}}]{Lebouteiller2017}
{Lebouteiller}, V., {P{\'e}quignot}, D., {Cormier}, D., {et~al.} 2017, \aap,
  602, A45

\bibitem[{{Lehmann} {et~al.}(2005){Lehmann}, {Becker}, {Fabrika}, {Roth},
  {Miyaji}, {Afanasiev}, {Sholukhova}, {S{\'a}nchez}, {Greiner}, {Hasinger},
  {Costantini}, {Surkov}, \& {Burenkov}}]{Lehmann2005}
{Lehmann}, I., {Becker}, T., {Fabrika}, S., {et~al.} 2005, \aap, 431, 847

\bibitem[{{Lehmer} {et~al.}(2016){Lehmer}, {Basu-Zych}, {Mineo}, {Brandt},
  {Eufrasio}, {Fragos}, {Hornschemeier}, {Luo}, {Xue}, {Bauer}, {Gilfanov},
  {Ranalli}, {Schneider}, {Shemmer}, {Tozzi}, {Trump}, {Vignali}, {Wang},
  {Yukita}, \& {Zezas}}]{Lehmer2016}
{Lehmer}, B.~D., {Basu-Zych}, A.~R., {Mineo}, S., {et~al.} 2016, \apj, 825, 7

\bibitem[{{Lehmer} {et~al.}(2021){Lehmer}, {Eufrasio}, {Basu-Zych}, {Doore},
  {Fragos}, {Garofali}, {Kovlakas}, {Williams}, {Zezas}, \&
  {Santana-Silva}}]{Lehmer2021}
{Lehmer}, B.~D., {Eufrasio}, R.~T., {Basu-Zych}, A., {et~al.} 2021, \apj, 907,
  17

\bibitem[{{Liu} {et~al.}(2015){Liu}, {Bai}, {Wang}, {Justham}, {Lu}, {Gu},
  {Liu}, {di Stefano}, {Guo}, {Cabrera-Lavers}, {{\'A}lvarez}, {Cao}, \&
  {Kulkarni}}]{Liu2015}
{Liu}, J.-F., {Bai}, Y., {Wang}, S., {et~al.} 2015, \nat, 528, 108

\bibitem[{{L{\'o}pez} {et~al.}(2019){L{\'o}pez}, {Jonker}, {Heida}, {Torres},
  {Roberts}, {Walton}, {Moon}, \& {Harrison}}]{Lopez2019}
{L{\'o}pez}, K.~M., {Jonker}, P.~G., {Heida}, M., {et~al.} 2019, \mnras, 489,
  1249

\bibitem[{{Madau} \& {Fragos}(2017)}]{Madau2017}
{Madau}, P. \& {Fragos}, T. 2017, \apj, 840, 39

\bibitem[{{Maggi} {et~al.}(2011){Maggi}, {Hou}, \& {Pakull}}]{Maggi2011}
{Maggi}, P., {Hou}, X., \& {Pakull}, M. 2011, in The X-ray Universe 2011, ed.
  J.-U. {Ness} \& M.~{Ehle}, 247

\bibitem[{{Makishima} {et~al.}(2000){Makishima}, {Kubota}, {Mizuno}, {Ohnishi},
  {Tashiro}, {Aruga}, {Asai}, {Dotani}, {Mitsuda}, {Ueda}, {Uno}, {Yamaoka},
  {Ebisawa}, {Kohmura}, \& {Okada}}]{Makishima2000}
{Makishima}, K., {Kubota}, A., {Mizuno}, T., {et~al.} 2000, \apj, 535, 632

\bibitem[{Middleton {et~al.}(2011)Middleton, Roberts, Done, \&
  Jackson}]{Middleton2011}
Middleton, M.~J., Roberts, T.~P., Done, C., \& Jackson, F.~E. 2011, Monthly
  Notices of the Royal Astronomical Society, 411, 644

\bibitem[{{Miller} {et~al.}(2005){Miller}, {Mushotzky}, \& {Neff}}]{Miller2005}
{Miller}, N.~A., {Mushotzky}, R.~F., \& {Neff}, S.~G. 2005, \apjl, 623, L109

\bibitem[{{Mineo} {et~al.}(2014){Mineo}, {Gilfanov}, {Lehmer}, {Morrison}, \&
  {Sunyaev}}]{Mineo2014}
{Mineo}, S., {Gilfanov}, M., {Lehmer}, B.~D., {Morrison}, G.~E., \& {Sunyaev},
  R. 2014, \mnras, 437, 1698

\bibitem[{{Mineo} {et~al.}(2012){Mineo}, {Gilfanov}, \& {Sunyaev}}]{Mineo2012b}
{Mineo}, S., {Gilfanov}, M., \& {Sunyaev}, R. 2012, \mnras, 426, 1870

\bibitem[{{Mitsuda} {et~al.}(1984){Mitsuda}, {Inoue}, {Koyama}, {Makishima},
  {Matsuoka}, {Ogawara}, {Shibazaki}, {Suzuki}, {Tanaka}, \&
  {Hirano}}]{Mitsuda1984}
{Mitsuda}, K., {Inoue}, H., {Koyama}, K., {et~al.} 1984, \pasj, 36, 741

\bibitem[{{Moon} {et~al.}(2011){Moon}, {Harrison}, {Cenko}, \&
  {Shariff}}]{Moon2011Large-Highly-Io}
{Moon}, D.-S., {Harrison}, F.~A., {Cenko}, S.~B., \& {Shariff}, J.~A. 2011,
  \apjl, 731, L32

\bibitem[{{Nakajima} {et~al.}(2018){Nakajima}, {Schaerer}, {Le F{\`e}vre},
  {Amor{\'\i}n}, {Talia}, {Lemaux}, {Tasca}, {Vanzella}, {Zamorani},
  {Bardelli}, {Grazian}, {Guaita}, {Hathi}, {Pentericci}, \&
  {Zucca}}]{Nakajima2018}
{Nakajima}, K., {Schaerer}, D., {Le F{\`e}vre}, O., {et~al.} 2018, \aap, 612,
  A94

\bibitem[{{Nanayakkara} {et~al.}(2019){Nanayakkara}, {Brinchmann}, {Boogaard},
  {Bouwens}, {Cantalupo}, {Feltre}, {Kollatschny}, {Marino}, {Maseda},
  {Matthee}, {Paalvast}, {Richard}, \& {Verhamme}}]{Nanayakkara2019}
{Nanayakkara}, T., {Brinchmann}, J., {Boogaard}, L., {et~al.} 2019, \aap, 624,
  A89

\bibitem[{{Oskinova} {et~al.}(2019){Oskinova}, {Bik}, {Mas-Hesse}, {Hayes},
  {Adamo}, {{\"O}stlin}, {F{\"u}rst}, \& {Ot{\'\i}-Floranes}}]{Oskinova2019}
{Oskinova}, L.~M., {Bik}, A., {Mas-Hesse}, J.~M., {et~al.} 2019, \aap, 627, A63

\bibitem[{{Pakull} \& {Angebault}(1986)}]{Pakull1986}
{Pakull}, M.~W. \& {Angebault}, L.~P. 1986, \nat, 322, 511

\bibitem[{{Pakull} \& {Mirioni}(2002)}]{Pakull2002}
{Pakull}, M.~W. \& {Mirioni}, L. 2002, arXiv e-prints, astro

\bibitem[{{Pakull} \& {Mirioni}(2003)}]{Pakull2003}
{Pakull}, M.~W. \& {Mirioni}, L. 2003, in Revista Mexicana de Astronomia y
  Astrofisica Conference Series, Vol.~15, Revista Mexicana de Astronomia y
  Astrofisica Conference Series, ed. J.~{Arthur} \& W.~J. {Henney}, 197--199

\bibitem[{{Pakull} {et~al.}(2010){Pakull}, {Soria}, \& {Motch}}]{Pakull2010}
{Pakull}, M.~W., {Soria}, R., \& {Motch}, C. 2010, \nat, 466, 209

\bibitem[{{P{\'e}rez-Montero} {et~al.}(2020){P{\'e}rez-Montero}, {Kehrig},
  {V{\'\i}lchez}, {Garc{\'\i}a-Benito}, {Duarte Puertas}, \&
  {Iglesias-P{\'a}ramo}}]{PerezMontero2020}
{P{\'e}rez-Montero}, E., {Kehrig}, C., {V{\'\i}lchez}, J.~M., {et~al.} 2020,
  arXiv e-prints, arXiv:2009.11076

\bibitem[{{Pintore} \& {Mereghetti}(2016)}]{Pintore2016}
{Pintore}, F. \& {Mereghetti}, S. 2016, \mnras, 460, 1033

\bibitem[{{Plat} {et~al.}(2019){Plat}, {Charlot}, {Bruzual}, {Feltre},
  {Vidal-Garc{\'\i}a}, {Morisset}, {Chevallard}, \& {Todt}}]{Plat_2019}
{Plat}, A., {Charlot}, S., {Bruzual}, G., {et~al.} 2019, \mnras, 490, 978

\bibitem[{{Poutanen} \& {Svensson}(1996)}]{Poutanen1996}
{Poutanen}, J. \& {Svensson}, R. 1996, \apj, 470, 249

\bibitem[{{Raiter} {et~al.}(2010){Raiter}, {Schaerer}, \&
  {Fosbury}}]{Raiter2010}
{Raiter}, A., {Schaerer}, D., \& {Fosbury}, R.~A.~E. 2010, \aap, 523, A64

\bibitem[{{Ramambason} {et~al.}(2020){Ramambason}, {Schaerer}, {Stasi{\'n}ska},
  {Izotov}, {Guseva}, {V{\'\i}lchez}, {Amor{\'\i}n}, \&
  {Morisset}}]{Ramambason2020}
{Ramambason}, L., {Schaerer}, D., {Stasi{\'n}ska}, G., {et~al.} 2020, \aap,
  644, A21

\bibitem[{{Ranalli} {et~al.}(2003){Ranalli}, {Comastri}, \&
  {Setti}}]{Ranalli2003}
{Ranalli}, P., {Comastri}, A., \& {Setti}, G. 2003, \aap, 399, 39

\bibitem[{{Rickards Vaught} {et~al.}(2021){Rickards Vaught}, {Sandstrom}, \&
  {Hunt}}]{Rickards2021}
{Rickards Vaught}, R.~J., {Sandstrom}, K.~M., \& {Hunt}, L.~K. 2021, arXiv
  e-prints, arXiv:2104.03931

\bibitem[{{Saxena} {et~al.}(2020{\natexlab{a}}){Saxena}, {Pentericci},
  {Mirabelli}, {Schaerer}, {Schneider}, {Cullen}, {Amorin}, {Bolzonella},
  {Bongiorno}, {Carnall}, {Castellano}, {Cucciati}, {Fontana}, {Fynbo},
  {Garilli}, {Gargiulo}, {Guaita}, {Hathi}, {Hutchison}, {Koekemoer}, {Marchi},
  {McLeod}, {McLure}, {Papovich}, {Pozzetti}, {Talia}, \&
  {Zamorani}}]{Saxena2020AA}
{Saxena}, A., {Pentericci}, L., {Mirabelli}, M., {et~al.} 2020{\natexlab{a}},
  \aap, 636, A47

\bibitem[{{Saxena} {et~al.}(2020{\natexlab{b}}){Saxena}, {Pentericci},
  {Schaerer}, {Schneider}, {Amorin}, {Bongiorno}, {Calabr{\`o}}, {Castellano},
  {Cimatti}, {Cullen}, {Fontana}, {Fynbo}, {Hathi}, {McLeod}, {Talia}, \&
  {Zamorani}}]{Saxena2020MNRAS}
{Saxena}, A., {Pentericci}, L., {Schaerer}, D., {et~al.} 2020{\natexlab{b}},
  \mnras, 496, 3796

\bibitem[{{Schaerer}(1996)}]{Schaerer1996}
{Schaerer}, D. 1996, \apjl, 467, L17

\bibitem[{{Schaerer} {et~al.}(2019){Schaerer}, {Fragos}, \&
  {Izotov}}]{Schaerer2019}
{Schaerer}, D., {Fragos}, T., \& {Izotov}, Y.~I. 2019, \aap, 622, L10

\bibitem[{Schaerer {et~al.}(2020)Schaerer, Izotov, \& Fragos}]{Schaerer2019b}
Schaerer, D., Izotov, Y., \& Fragos, T. 2020, Proceedings of the International
  Astronomical Union, 15, 79

\bibitem[{{Schaerer} \& {Stasi{\'n}ska}(1999)}]{Schaerer1999}
{Schaerer}, D. \& {Stasi{\'n}ska}, G. 1999, \aap, 345, L17

\bibitem[{{Schmidt} {et~al.}(2017){Schmidt}, {Huang}, {Treu}, {Hoag}, {Brada{\v
  c}}, {Henry}, {Jones}, {Mason}, {Malkan}, {Morishita}, {Pentericci},
  {Trenti}, {Vulcani}, \& {Wang}}]{Schmidt2017}
{Schmidt}, K.~B., {Huang}, K.-H., {Treu}, T., {et~al.} 2017, \apj, 839, 17

\bibitem[{{Schmidt} {et~al.}(2021){Schmidt}, {Kerutt}, {Wisotzki}, {Urrutia},
  {Feltre}, {Maseda}, {Nanayakkara}, {Bacon}, {Boogaard}, {Conseil}, {Contini},
  {Herenz}, {Kollatschny}, {Krumpe}, {Leclercq}, {Mahler}, {Matthee},
  {Mauerhofer}, {Richard}, \& {Schaye}}]{Schmidt2021}
{Schmidt}, K.~B., {Kerutt}, J., {Wisotzki}, L., {et~al.} 2021, arXiv e-prints,
  arXiv:2108.01713

\bibitem[{{Senchyna} {et~al.}(2020){Senchyna}, {Stark}, {Mirocha}, {Reines},
  {Charlot}, {Jones}, \& {Mulchaey}}]{Senchyna2020}
{Senchyna}, P., {Stark}, D.~P., {Mirocha}, J., {et~al.} 2020, \mnras, 494, 941

\bibitem[{{Senchyna} {et~al.}(2017){Senchyna}, {Stark}, {Vidal-Garc{\'\i}a},
  {Chevallard}, {Charlot}, {Mainali}, {Jones}, {Wofford}, {Feltre}, \&
  {Gutkin}}]{Senchyna2017}
{Senchyna}, P., {Stark}, D.~P., {Vidal-Garc{\'\i}a}, A., {et~al.} 2017, \mnras,
  472, 2608

\bibitem[{{Shirazi} \& {Brinchmann}(2012)}]{Shirazi2012}
{Shirazi}, M. \& {Brinchmann}, J. 2012, \mnras, 421, 1043

\bibitem[{{Siana} {et~al.}(2015){Siana}, {Shapley}, {Kulas}, {Nestor},
  {Steidel}, {Teplitz}, {Alavi}, {Brown}, {Conselice}, {Ferguson}, {Dickinson},
  {Giavalisco}, {Colbert}, {Bridge}, {Gardner}, \& {de Mello}}]{Siana2015}
{Siana}, B., {Shapley}, A.~E., {Kulas}, K.~R., {et~al.} 2015, \apj, 804, 17

\bibitem[{{Sobral} {et~al.}(2015){Sobral}, {Matthee}, {Darvish}, {Schaerer},
  {Mobasher}, {R{\"o}ttgering}, {Santos}, \& {Hemmati}}]{Sobral2015}
{Sobral}, D., {Matthee}, J., {Darvish}, B., {et~al.} 2015, \apj, 808, 139

\bibitem[{{Stanway} \& {Eldridge}(2019)}]{Stanway2019}
{Stanway}, E.~R. \& {Eldridge}, J.~J. 2019, \aap, 621, A105

\bibitem[{{Stark}(2016)}]{Stark2016}
{Stark}, D.~P. 2016, \araa, 54, 761

\bibitem[{{Stark} {et~al.}(2015){Stark}, {Walth}, {Charlot}, {Cl{\'e}ment},
  {Feltre}, {Gutkin}, {Richard}, {Mainali}, {Robertson}, {Siana}, {Tang}, \&
  {Schenker}}]{Stark2015}
{Stark}, D.~P., {Walth}, G., {Charlot}, S., {et~al.} 2015, \mnras, 454, 1393

\bibitem[{{Stasi{\'n}ska} {et~al.}(2015){Stasi{\'n}ska}, {Izotov}, {Morisset},
  \& {Guseva}}]{Stasinska2015}
{Stasi{\'n}ska}, G., {Izotov}, Y., {Morisset}, C., \& {Guseva}, N. 2015, \aap,
  576, A83

\bibitem[{{Stasi{\'n}ska} \& {Tylenda}(1986)}]{Stasinska1986}
{Stasi{\'n}ska}, G. \& {Tylenda}, R. 1986, \aap, 155, 137

\bibitem[{Steidel {et~al.}(2018)Steidel, Bogosavljevi{\'c}, Shapley, Reddy,
  Rudie, Pettini, Trainor, \& Strom}]{Steidel2018}
Steidel, C.~C., Bogosavljevi{\'c}, M., Shapley, A.~E., {et~al.} 2018, The
  Astrophysical Journal, 869, 123

\bibitem[{{Steidel} {et~al.}(2016){Steidel}, {Strom}, {Pettini}, {Rudie},
  {Reddy}, \& {Trainor}}]{Steidel2016}
{Steidel}, C.~C., {Strom}, A.~L., {Pettini}, M., {et~al.} 2016, \apj, 826, 159

\bibitem[{{Sturm} {et~al.}(2002){Sturm}, {Lutz}, {Verma}, {Netzer},
  {Sternberg}, {Moorwood}, {Oliva}, \& {Genzel}}]{Sturm2002}
{Sturm}, E., {Lutz}, D., {Verma}, A., {et~al.} 2002, \aap, 393, 821

\bibitem[{{Sutton} {et~al.}(2013{\natexlab{a}}){Sutton}, {Roberts},
  {Gladstone}, {Farrell}, {Reilly}, {Goad}, \& {Gehrels}}]{Sutton2013a}
{Sutton}, A.~D., {Roberts}, T.~P., {Gladstone}, J.~C., {et~al.}
  2013{\natexlab{a}}, \mnras, 434, 1702

\bibitem[{{Sutton} {et~al.}(2013{\natexlab{b}}){Sutton}, {Roberts}, \&
  {Middleton}}]{Sutton2013b}
{Sutton}, A.~D., {Roberts}, T.~P., \& {Middleton}, M.~J. 2013{\natexlab{b}},
  \mnras, 435, 1758

\bibitem[{{Svoboda} {et~al.}(2019){Svoboda}, {Douna}, {Orlitov{\'a}}, \&
  {Ehle}}]{Svoboda2019}
{Svoboda}, J., {Douna}, V., {Orlitov{\'a}}, I., \& {Ehle}, M. 2019, \apj, 880,
  144

\bibitem[{{Sz{\'e}csi} {et~al.}(2015){Sz{\'e}csi}, {Langer}, {Yoon}, {Sanyal},
  {de Mink}, {Evans}, \& {Dermine}}]{Szecsi2015}
{Sz{\'e}csi}, D., {Langer}, N., {Yoon}, S.-C., {et~al.} 2015, \aap, 581, A15

\bibitem[{{Tao} {et~al.}(2012){Tao}, {Kaaret}, {Feng}, \&
  {Gris{\'e}}}]{Tao2012}
{Tao}, L., {Kaaret}, P., {Feng}, H., \& {Gris{\'e}}, F. 2012, \apj, 750, 110

\bibitem[{{Thuan} \& {Izotov}(2005)}]{Thuan2005}
{Thuan}, T.~X. \& {Izotov}, Y.~I. 2005, \apjs, 161, 240

\bibitem[{{Titarchuk} {et~al.}(1997){Titarchuk}, {Mastichiadis}, \&
  {Kylafis}}]{Titarchuk1997}
{Titarchuk}, L., {Mastichiadis}, A., \& {Kylafis}, N.~D. 1997, \apj, 487, 834

\bibitem[{{Urquhart} {et~al.}(2018){Urquhart}, {Soria}, {Johnston}, {Pakull},
  {Motch}, {Schwope}, {Miller-Jones}, \& {Anderson}}]{Urquhart2018}
{Urquhart}, R., {Soria}, R., {Johnston}, H.~M., {et~al.} 2018, \mnras, 475,
  3561

\bibitem[{{Urquhart} {et~al.}(2019){Urquhart}, {Soria}, {Pakull},
  {Miller-Jones}, {Anderson}, {Plotkin}, {Motch}, {Maccarone}, {McLeod}, \&
  {Scaringi}}]{Urquhart2019}
{Urquhart}, R., {Soria}, R., {Pakull}, M.~W., {et~al.} 2019, \mnras, 482, 2389

\bibitem[{{Vinokurov} {et~al.}(2018){Vinokurov}, {Fabrika}, \&
  {Atapin}}]{Vinokurov2018}
{Vinokurov}, A., {Fabrika}, S., \& {Atapin}, K. 2018, \apj, 854, 176

\bibitem[{{Walton} {et~al.}(2015){Walton}, {Middleton}, {Rana}, {Miller},
  {Harrison}, {Fabian}, {Bachetti}, {Barret}, {Boggs}, {Christensen}, {Craig},
  {Fuerst}, {Grefenstette}, {Hailey}, {Madsen}, {Stern}, \&
  {Zhang}}]{Walton2015}
{Walton}, D.~J., {Middleton}, M.~J., {Rana}, V., {et~al.} 2015, \apj, 806, 65

\end{thebibliography}
\bibliographystyle{aa}

\newpage
\begin{appendix}
\section{Comparison of Cloudy predictions with Case B}

\begin{figure}[h]
   \centering
   \includegraphics[width=1\columnwidth]{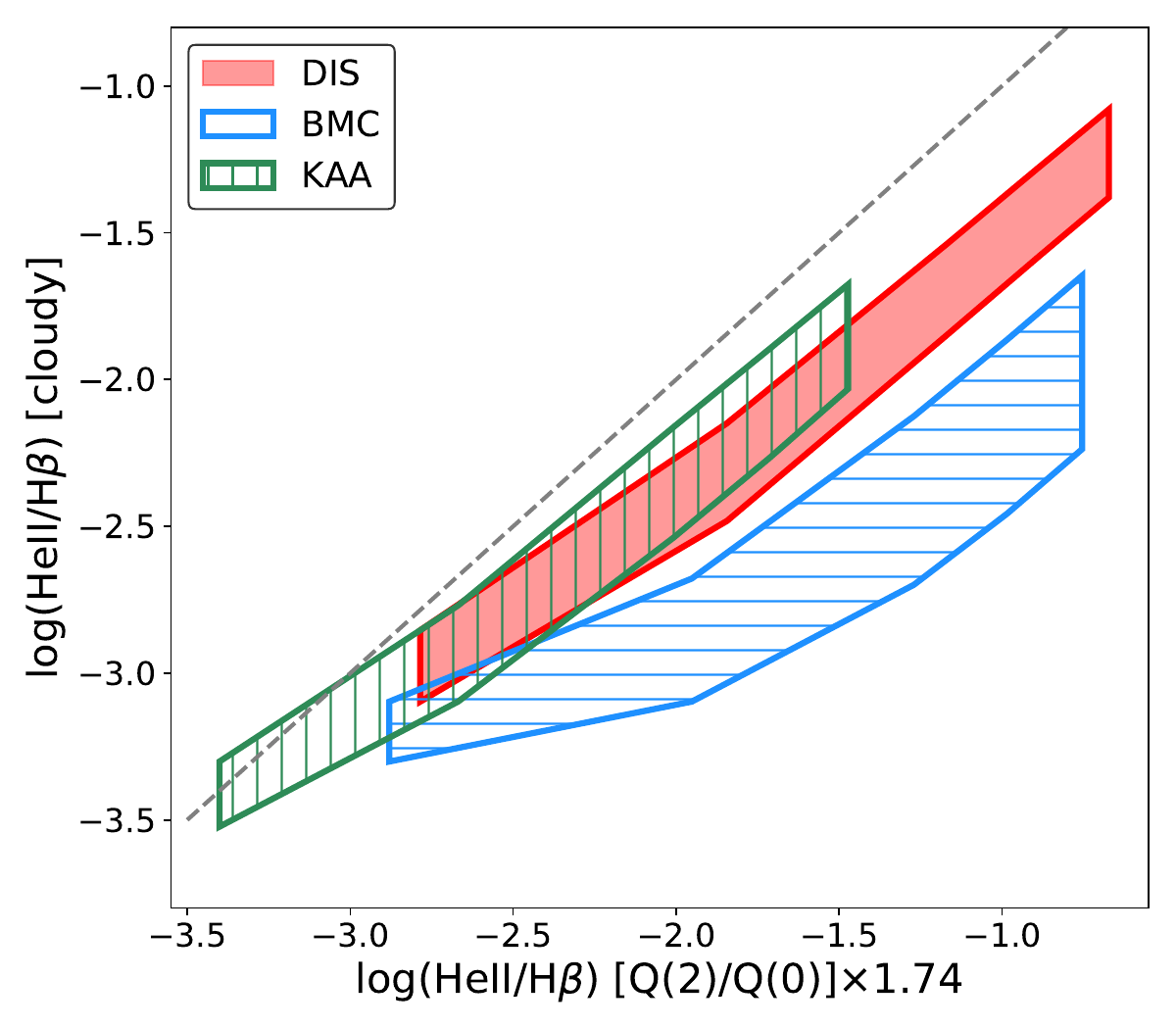}
   \caption{Percentages of He$\rm{II}$ in relation to H$\beta$. The X-axis shows the analytical predictions based on the input spectra, where the factor 1.74 corresponds to typical nebular conditions.  The dashed line shows a one-to-one relation between the predictions and the outputs. The upper and lower boundary for each model correspond to $\log U = -1.5$ and $\log U = -3.5$, respectively.} 

    \label{fig_Qs}%
\end{figure}

To first order, assuming simple case B recombination line emissivities, the relative line intensities of \heii/\hb\ is given by
\begin{equation}
    I(4686)/I(\hb) = A \times \frac{Q_2}{Q_0},
    \label{eq_q2q0}
\end{equation}
where $Q_0$ and $Q_2$ are the ionising photon fluxes for H and \hep\ respectively (i.e.\ above 13.6 and 54 eV), and $A \simeq 1.74$ for typical nebular conditions \citep{Stasinska2015}.
In Fig.\ \ref{fig_Qs}, we compare the line ratio calculated from this simple analytic expression with the predictions from our Cloudy model grid.
Overall the scaling is as expected, and the models with high ionisation parameter agree well with Eq.\ \ref{eq_q2q0}, except for very low values of 
\Heii/\hb $\la 3 \times 10^{-3}$, which is below the current detection limit. For models with low $U$, we find a decrease of \heii/\hb, by up to a factor $\sim 2-3$, 
which we attribute to the competition between H and He, discussed in earlier studies \citep[e.g.][]{Stasinska1986,Raiter2010}, and which cannot be captured by simple
Case B theory. 
We also note that for the hardest SED (between 54 eV, the ionisation potential of He$^+$, and the X-ray domain) Cloudy predicts a lower \heii/\hb\ intensity than naively expected from simple ionising-photon counting.
This is most likely due to the fact only a fraction of photons with energy $>54$ eV have energies sufficiently close to the ionisation potential to be efficiently absorbed by He$^+$, whose photoionisation cross section decreases with $\nu^{-3}$.

\end{appendix}

\end{document}